\documentclass[twocolumn]{aastex63}

\newcommand{\todo}{\ifmmode \text{\color{purple}\Huge{\(\bullet\)}} \else {\color{purple}{\Huge$\bullet$}}\fi}
\newcommand{\finish}{\ifmmode \text{\color{blue}\Huge{\(\bullet\)}} \else {\color{blue}{\Huge$\bullet$}}\fi}

\newcommand{\NH}{N_\mathrm{H}}
\newcommand{\NHunit}{N_\mathrm{H}/\mathrm{cm}^{-2}}

\newcommand{\msun}{M_{\odot}}
\newcommand{\mbh}{M_\mathrm{BH}}

\newcommand{\av}{A_\mathrm{V}}
\newcommand{\R}{\mathcal{R_{\rm rest}}}

\newcommand{\lamrest}{\lambda_\mathrm{rest}}

\newcommand{\mgtwo}{Mg\,\textsc{ii}$\lambda2800$}
\newcommand{\oiii}{[O\,\textsc{iii}]$\lambda5008$}

\newcommand{\mgtwodou}{Mg\,\textsc{ii}$\lambda\lambda2796,2803$\xspace}

\newcommand{\feii}{Fe\,\textsc{ii}\xspace}

\newcommand{\kms}{\mathrm{km\ s^{-1}}}
\newcommand{\ergs}{\mathrm{erg\ s^{-1}}}

\newcommand{\tnr}[2]{{#1}_\mathrm{#2}\xspace}
\newcommand{\targetone}{JADES1180}
\newcommand{\targettwo}{GTO1215}
\newcommand{\sanfit}{\mathtt{S^{3}Fit}}

\definecolor{darkgoldenrod}{HTML}{B8860B}

\usepackage{natbib}
\usepackage{amsmath}
\usepackage{multirow}
\usepackage{ifthen}
\usepackage{tabularx}
\usepackage{xspace}

\received{\today}

\submitjournal{ApJL}

\shorttitle{JWST BBQs}
\shortauthors{Boonmee et al.}

\begin{document}

\title{
Hotter than Little Red Dots: Two Quasars with $T\sim10^4$~K blackbody-like continua at $z\sim4$
}

\correspondingauthor{Kristiphong Boonmee}
\email{kris@akane.waseda.jp,k.ichikawa@astr.tohoku.ac.jp}

\author[0009-0001-2570-2127]{Kristiphong Boonmee}
\affiliation{
Department of Pure and Applied Physics, Graduate School of Advanced Science and Engineering, Faculty of Science and Engineering, Waseda University, 3-4-1,
Okubo, Shinjuku, Tokyo 169-8555, Japan}

\author[0000-0002-4377-903X]{Kohei Ichikawa}
\affiliation{Frontier Research Institute for Interdisciplinary Sciences, Tohoku University, Sendai, Miyagi 980-8578, Japan}
\affiliation{Astronomical Institute, Tohoku University, Aramaki, Aoba-ku, Sendai, Miyagi 980-8578, Japan}

\author[0009-0001-3910-2288]{Yuxing Zhong}
\affiliation{Frontier Research Institute for Interdisciplinary Sciences, Tohoku University, Sendai, Miyagi 980-8578, Japan}

\author[0000-0003-2682-473X]{Xiaoyang Chen}
\affiliation{Frontier Research Institute for Interdisciplinary Sciences, Tohoku University, Sendai, Miyagi 980-8578, Japan}

\author[0000-0002-9754-3081]{Satoshi Yamada}
\affiliation{Frontier Research Institute for Interdisciplinary Sciences, Tohoku University, Sendai, Miyagi 980-8578, Japan}
\affiliation{Astronomical Institute, Tohoku University, Aramaki, Aoba-ku, Sendai, Miyagi 980-8578, Japan}
\affiliation{Department of Astronomy, University of Geneva, Ch.d’Ecogia 16, 1290, Versoix, Switzerland.}

\author[0000-0002-7779-8677]{Akio K. Inoue}
\affiliation{
Department of Pure and Applied Physics, Graduate School of Advanced Science and Engineering, Faculty of Science and Engineering, Waseda University, 3-4-1,
Okubo, Shinjuku, Tokyo 169-8555, Japan}
\affiliation{Waseda Research Institute of Science and Engineering, Faculty of Science and Engineering, Waseda University, 3-4-1 Okubo, Shinjuku, Tokyo 169-8555, Japan}

\author[0009-0001-9170-3363]{Jialai Wang}
\affiliation{Department of Astronomy, University of Science and Technology of China, Hefei 230026, China}
\affiliation{School of Astronomy and Space Science, University of Science and Technology of China, Hefei 230026, China}
\affiliation{Kavli Institute for the Physics and Mathematics of the Universe (Kavli IPMU, WPI), The University of Tokyo Institutes for Advanced Study (UTIAS), The University of Tokyo, Kashiwa, Chiba 277-8583, Japan}
\affiliation{Center for Data-Driven Discovery, Kavli IPMU (WPI), UTIAS, The University of Tokyo, Kashiwa, Chiba 277-8583, Japan}

\author[0009-0002-4201-7727]{Youwen Kong}
\affiliation{Institute of Astronomy, Graduate School of Science, The University of Tokyo, 2-21-1 Osawa, Mitaka, Tokyo, 181-0015 Japan}

\author[0009-0007-6567-4240]{Sakiko Obuchi}
\affiliation{
Department of Physics, School of Advanced Science and Engineering, Faculty of Science and Engineering, Waseda University, 3-4-1,
Okubo, Shinjuku, Tokyo 169-8555, Japan}

\author[0009-0001-0366-9266]{John Schneider}
\affiliation{
Department of Physics and Astronomy, The University of Texas at San Antonio, 1 UTSA Circle, San Antonio, TX 78249, USA}

\author[0000-0003-2213-7983]{Bovornpratch Vijarnwannaluk}
\affiliation{Academia Sinica Institute of Astronomy and Astrophysics,
11F of Astronomy-Mathematics Building, AS/NTU, No.1, Section 4,
Roosevelt Road, Taipei 10617, Taiwan}

\begin{abstract}
Gas envelope models of Little Red Dots (LRDs) predict a transition
from cool ($T\sim5000$~K), gas-enshrouded accretion to an exposed quasar,
but observational candidates remain scarce. We identify two blackbody quasars (BBQs), \targetone\ at $z=3.711$ and \targettwo\ at $z=4.552$, among 3,553 sources with archival NIRSpec/PRISM spectra.  Both show extremely red rest-frame UV ($\beta_\mathrm{UV}>2.0$) but blue optical continua ($\beta_\mathrm{opt}<-0.75$), where $f_\lambda\propto\lambda^\beta$.
Their $\Lambda$-shaped UV--optical continua peak at $\lambda\approx3000$--$4000$~\AA\ and are not reproduced by a quasar template reddened with an SMC-like extinction curve.
Under the gas envelope interpretation, UV-to-mid-infrared spectral energy distribution
fitting indicates $T_\mathrm{env}\approx7600$--$10300$~K and $L_\mathrm{env}\simeq (3-15)\times10^{45}~\ergs$, placing both sources just above the typical LRD ranges in temperature and luminosity.
Both BBQs show hot-dust emission with $T_\mathrm{dust}\simeq1400$--$1550$~K
and X-ray emission with $L_\mathrm{2-10keV}\simeq(1.3$--$4.5)\times10^{44}\ \ergs$.
The gas absorption is $\NH < 1.5\times10^{23}$~cm$^{-2}$, indicating Compton-thin absorption sightlines.
Both also exhibit far-infrared and millimeter detections, suggesting an additional cooler dust component, potentially associated with their host galaxies.
These properties are consistent with a transitional phase in which a hotter gaseous envelope partially exposes the central engine.
\end{abstract}


\section{Introduction}\label{sec:intro}

One of the fundamental questions on supermassive black holes (SMBHs) 
is how their mass assembly has been achieved from the first BH formation \citep{ina20} to the current values \citep{kor13}.
James Webb Space Telescope (JWST) has discovered a new class of active galactic nuclei (AGN)
called little red dots \citep[LRDs; ][]{kov23,mat24}, exhibiting compact morphologies, a ``V''-shaped rest-UV to optical spectral energy distribution (SED), broad Balmer emission lines \citep[$>80$\%;][]{koc25,hvi25} and frequently accompanied by narrow absorption features \citep[$>30$\%;][]{juo24,lin24}.
However, most LRDs are different from typical type-1 AGNs, such as the scarcity of UV/optical variability \citep{kok24a,zha25a,zha25b}, the absence of X-ray, mid-infrared (MIR), and radio emissions \citep{ana24,yue24,wil24,set25}.

Recent studies raised the scenario that the central SMBHs are covered by
Compton-thick ($\NH \gg 10^{24}$~cm$^{-2}$) dense gas 
\citep{kid25,liu2025_lrd_balmer,beg26,nandal26,wendy26,Santarelli26},
and the energy balance naturally realizes the stellar-like blackbody or modified blackbody spectrum with $T\sim5000$--$7000$~K.
This gas configuration can also describe the unique properties of LRDs, such as suppression of radio and X-ray emission, redder optical continuum, and compact morphology \citep{ina26}.

LRDs also show a distinct difference in cosmic evolution from that of the luminous quasars.
LRDs are abundant in the early universe, with a peak around $z\sim7$ \citep{tan25} reaching $\sim10^{-4}$~Mpc$^{-3}$ \citep{ina24,ina25c}, but their number density declines rapidly at $z<4$ \citep{ma25a} and possibly reaching $\sim 10^{-9}$~Mpc$^{-3}$ at $z<0.5$ \citep{lin26,par26}.
Recent observations further suggest a sharp upper luminosity boundary at
$L_\mathrm{bol}\approx10^{45}$~erg~s$^{-1}$ \citep{ma25b}, corresponding
to $\mbh \approx 10^{7} M_\odot$ assuming Eddington-limited accretion.
This mass limit is also consistent with the maximum $\mbh$ expected
from the BH mass assembly on the Salpeter timescale, which is comparable
to the inferred LRD lifetime of $\sim 3 \times 10^{7}$~yrs \citep{kid25}.
These findings suggest that LRDs may trace a short-lived phase of a first generation of rapid SMBH growth at $z>4$. However, the fate of this population at later cosmic epochs remains poorly understood because relatively few LRD-like systems have been identified at $z<4$, except several reports at $z<1$ \citep{lin25,lin26,chen25_lrd,par26}. 

In the gas-envelope framework, once the gas reservoir is depleted, the envelope is expected to disperse and its photospheric temperature to increase, revealing the accretion disk \citep{hos12,hos13}.
This suggests that such a ``transition'' stage should exist between the LRDs and a classical unobscured quasar.
 Several reports of X-ray and/or radio-detected LRDs provide tentative observational evidence for such a transition system, although their inferred envelope temperatures remain relatively low  \citep[$T\sim5000$–7000 K;][]{fu25,hvi25}.

Recently, \cite{zho26} reported the discovery of an interesting AGN at $z=1.715$. The source shows broad \mgtwo \ and H$\alpha$ emission lines and its rest-frame UV-optical continuum shows $\Lambda$-shaped continuum around $\lamrest\approx2500$~\AA, which is well described by a blackbody spectrum with $T\sim10^4$~K, a factor of two hotter than that of typical LRDs \citep{deg25}. In addition, integrated luminosity of $L_\mathrm{BB}\approx10^{46}$~erg~s$^{-1}$ reaches as one of the most luminous level as LRDs.
Since they are detected in both X-ray and radio frequencies, they are dubbed BlackBody QSO and Radio Source (BBQSORS).
The properties of BBQSORS suggest that they may represent a short-lived evolutionary stage in which
the dense envelope is dispersed and the central quasar is partially exposed.
However, the evolutionary connection between high-redshift LRDs and lower-redshift transitional systems such as BBQSORS at $z\sim2$ remains poorly constrained. 
Identifying BBQSORS-like systems at $z=3$--$5$, where most LRDs are expected to experience the transition to typical quasars, is critical for tracing the transformation of LRDs into the typical quasar population for establishing the cosmic evolution of envelope-dominated SMBH growth.

In this letter, we present the finding of two blackbody quasars (BBQs) with $T\sim10^4$~K at $z\sim3.7$ and $z\sim4.6$ from the JWST archive, which expanded the redshift range into the regime of abundant LRD populations at $z>3$.
We adopt the following cosmological parameters throughout this paper: 
$H_0 = 68$\,km\,s$^{-1}$\,Mpc$^{-1}$, $\Omega_\mathrm{M}=0.31$, and $\Omega_\Lambda=0.69$ \citep{Planck2020}.

\begin{figure}
    \centering
    \includegraphics[width=0.95\columnwidth]{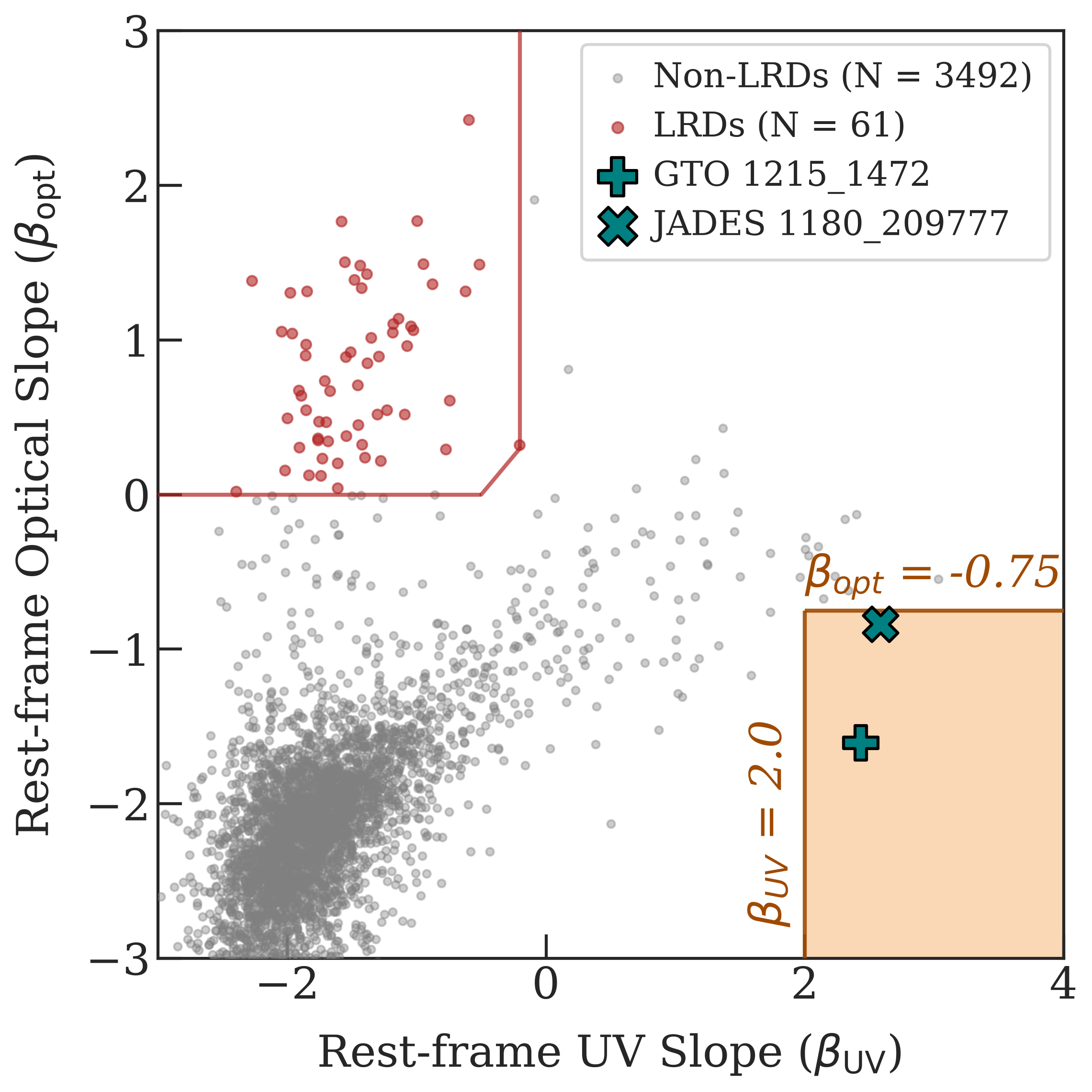}
    \caption{The rest-frame UV slope ($\beta_\mathrm{UV}$) against the rest-frame optical slope ($\beta_\mathrm{opt}$) for the 3,553 samples in \citet{boo26}. The red lines indicate the boundary criteria for identifying LRDs in \citet{deg25}. Red points indicate the sources classified as LRDs in \citet{boo26}, while the grey points shows the non-LRDs. The teal-colored plus and cross symbols illustrate the position of the two BBQ sources in the $\beta_{\mathrm{UV}}$–$\beta_{\mathrm{opt}}$ plane that are chosen for this study, based on the orange shading and boundaries.}
    \label{fig:betabeta}
\end{figure}

\section{Sample selection and Spectral Fitting}\label{sec:data_reduction}

\begin{figure*}[tp!]
\begin{center}
\includegraphics[width=0.45\textwidth]{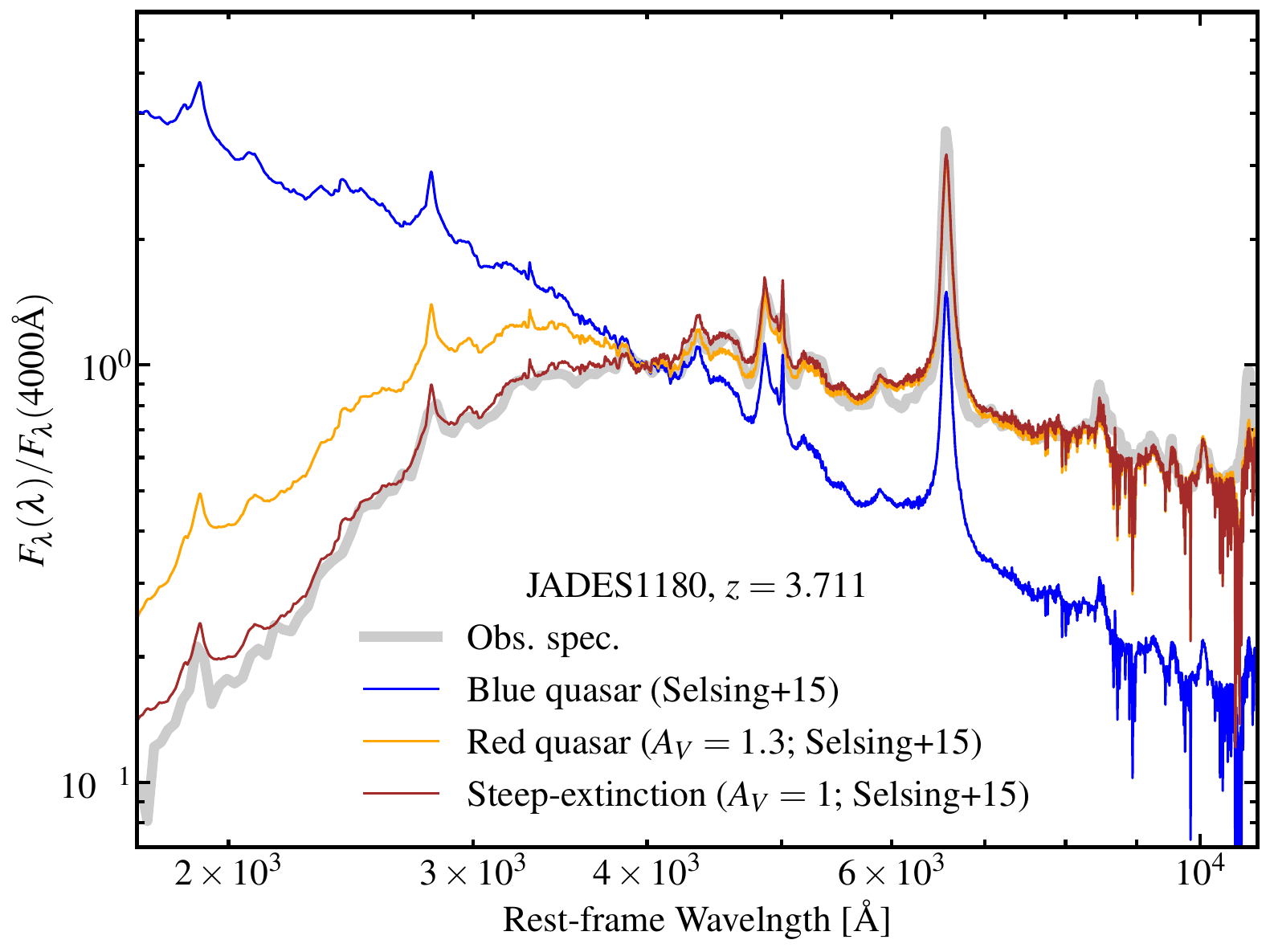}~
\includegraphics[width=0.45\textwidth]{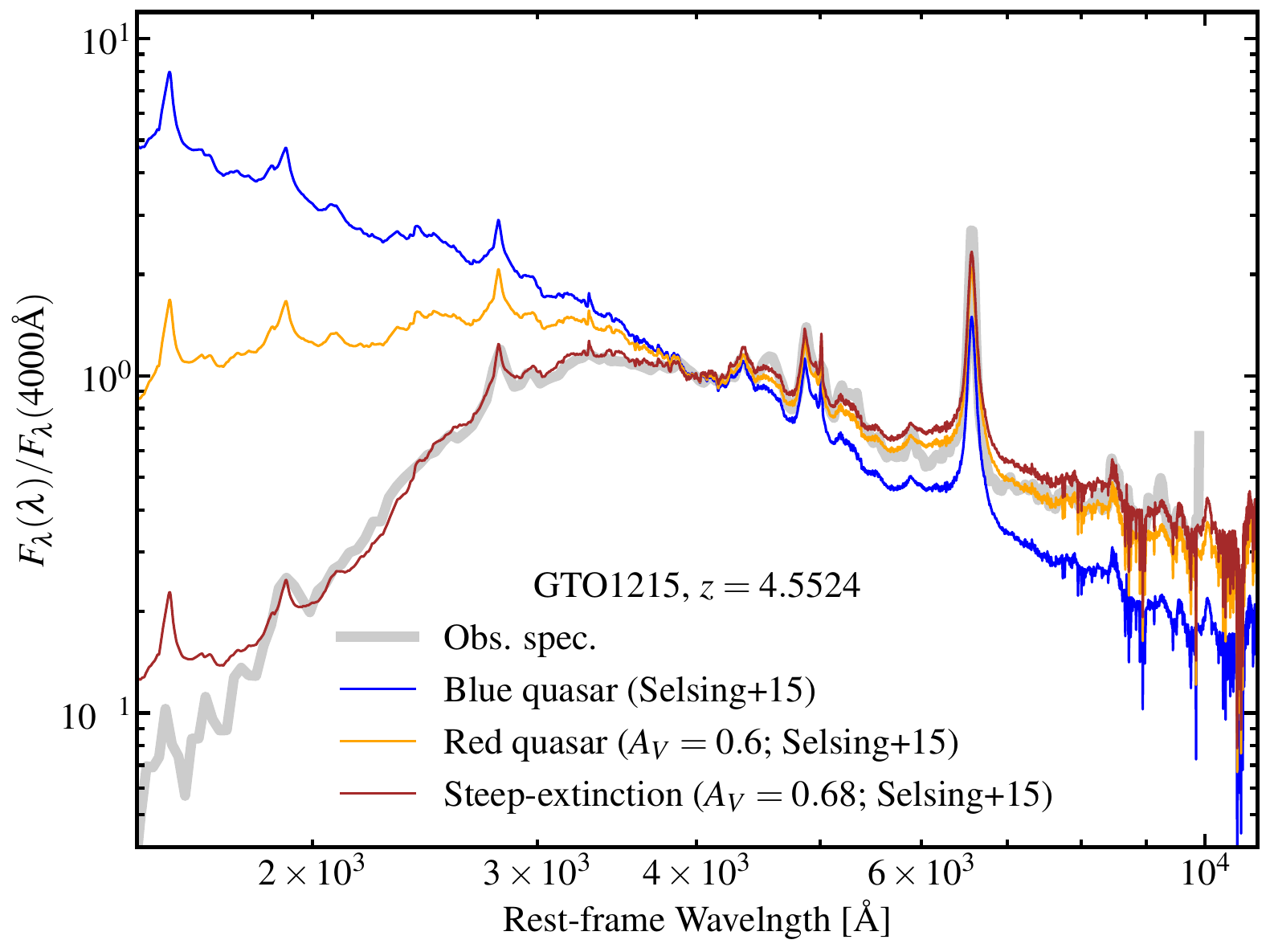}\\
\caption{
Comparison of the spectra among the two BBQs, blue quasar, and red quasars. 
The rest-frame spectra \targetone\ (left) and for \targettwo\ (right) are shown with gray shaded regions.
The blue lines show the blue quasar template of \cite{sel16}.
The orange lines represent the red quasar template (dust-reddened version of the blue quasar) assuming an SMC-like extinction curve \cite{Gordon2024_dust} with $\av = 1.3$~mag for \targetone, and $\av = 0.6$~mag for \targettwo, respectively.
The brown lines represent the red quasar template assuming steeper extinction curves (grain mass distribution skewed towards small grains) with $\av = 1.0$~mag for \targetone, and $\av = 0.68$~mag for \targettwo, respectively.
All spectra and SEDs are normalized at rest-frame $\lambda_\mathrm{rest}=4000$~\AA.
}
\label{fig:dustQSO}
\end{center}
\end{figure*}

We selected blackbody quasars (BBQs) with characteristic $\Lambda$-shaped continuum using
 archival JWST/PRISM spectroscopic data covering $6000~\mathrm{\AA} < \lambda < 55000$~$\mathrm{\AA}$ \citep{pol26} from the Dawn JWST Archive \citep[DJA;][]{hei24,deg25a}\footnote{https://dawn-cph.github.io/dja/spectroscopy/nirspec/}.
Using the latest public release (version 4.4), we selected all spectroscopically confirmed sources within the redshift range $3.0 \leq z_{\rm spec} \leq 8.1$, ensuring that for each object, the rest-frame wavelength range $1500~\mathrm{\AA} < \lambda_{\rm rest} < 6000~\mathrm{\AA}$ is covered by all sources. This selection left 7,904 highest quality (\texttt{grade} = 3) sources.

To investigate the continuum shape, we modeled the rest-frame UV and optical spectra with a simple power-law, $f_\lambda = \alpha \times\lamrest^{\beta_{\lambda}}$.
The UV and optical slopes were measured in $1500~\mathrm{\AA} < \lamrest < 3600~\mathrm{\AA}$ and  $4000~\mathrm{\AA} < \lamrest < 6000~\mathrm{\AA}$, respectively. 
For each source, we required $>10$ spectral data points in both intervals after masking the wavelength range within $100$~\AA\ of strong emission lines (e.g. H$\beta$, \oiii, \mgtwo). 
We also required a median continuum signal-to-noise ratio of $\mathrm{S/N}>2$ in both wavelength ranges. 
To ensure robust continuum measurements, we chose sources with uncertainties of $\sigma(\beta_\mathrm{UV})<1$ and $\sigma(\beta_\mathrm{opt})<1$. 
After applying all selection criteria and removing duplicate sources within a radius of 0.15$''$, the final sample consists of 3,553 unique sources.

\begin{figure*}[tp!]
\begin{center}
\includegraphics[width=0.5\textwidth]{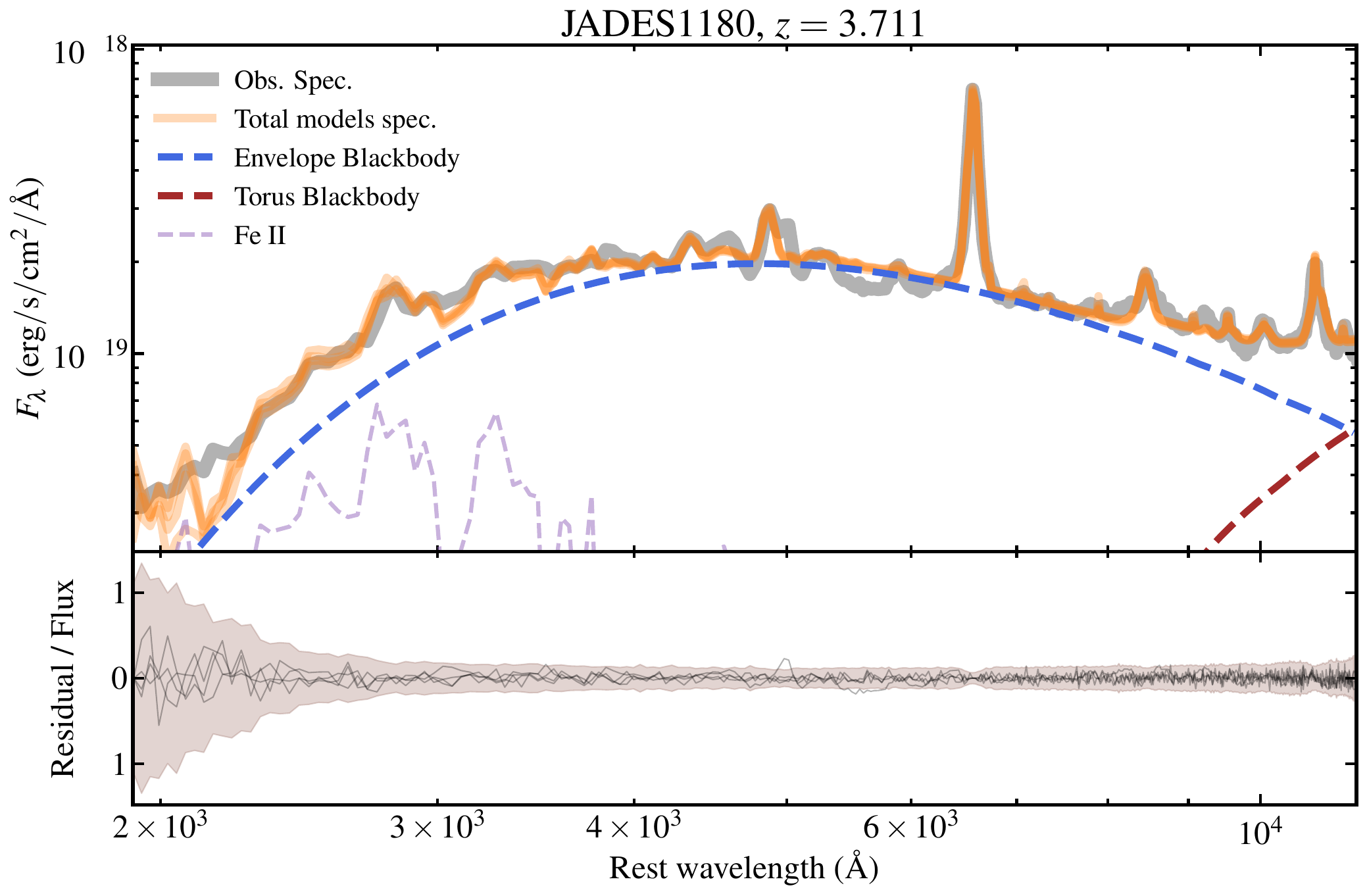}~
\includegraphics[width=0.5\textwidth]{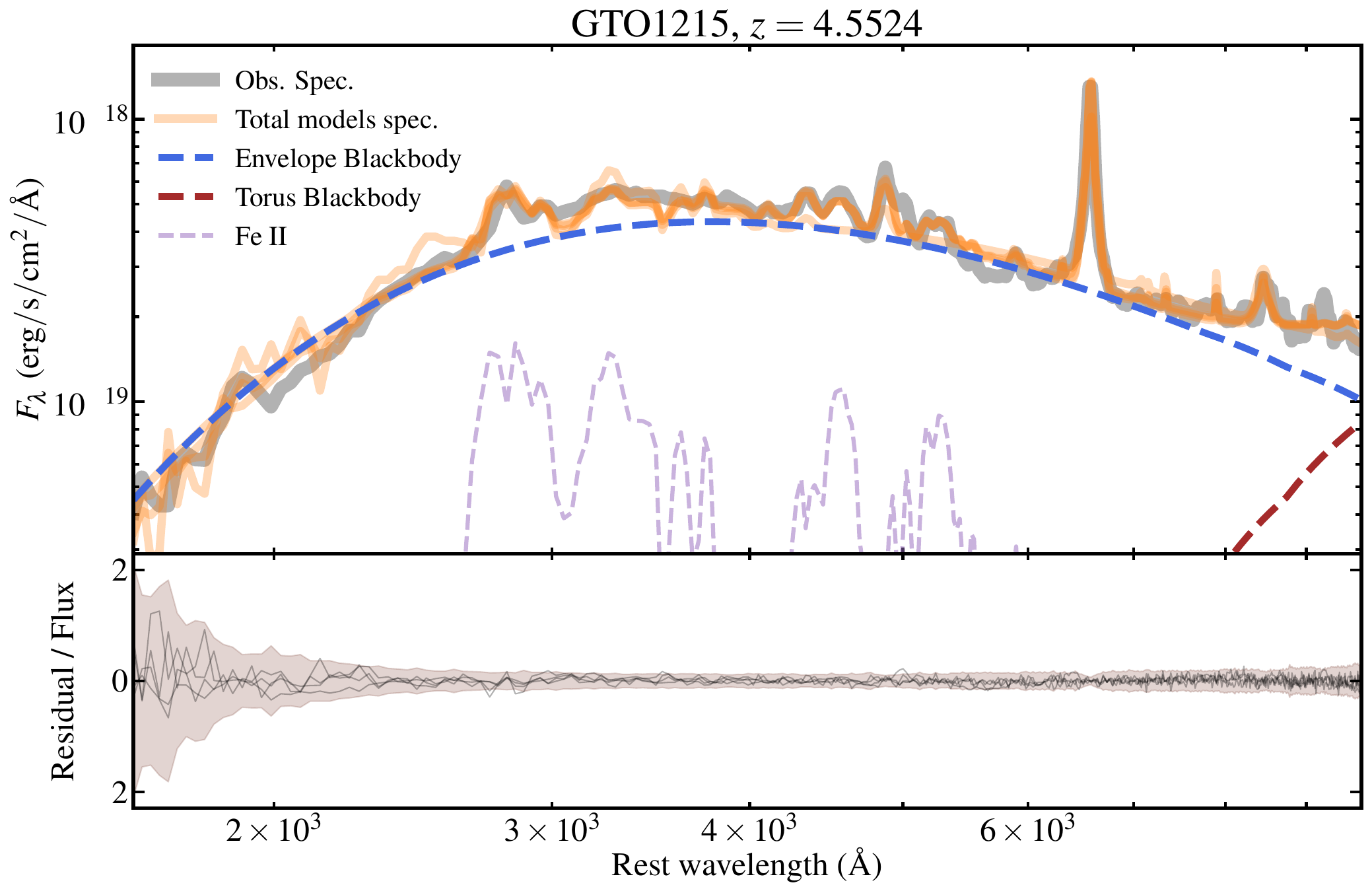}\\
\includegraphics[width=0.5\textwidth]{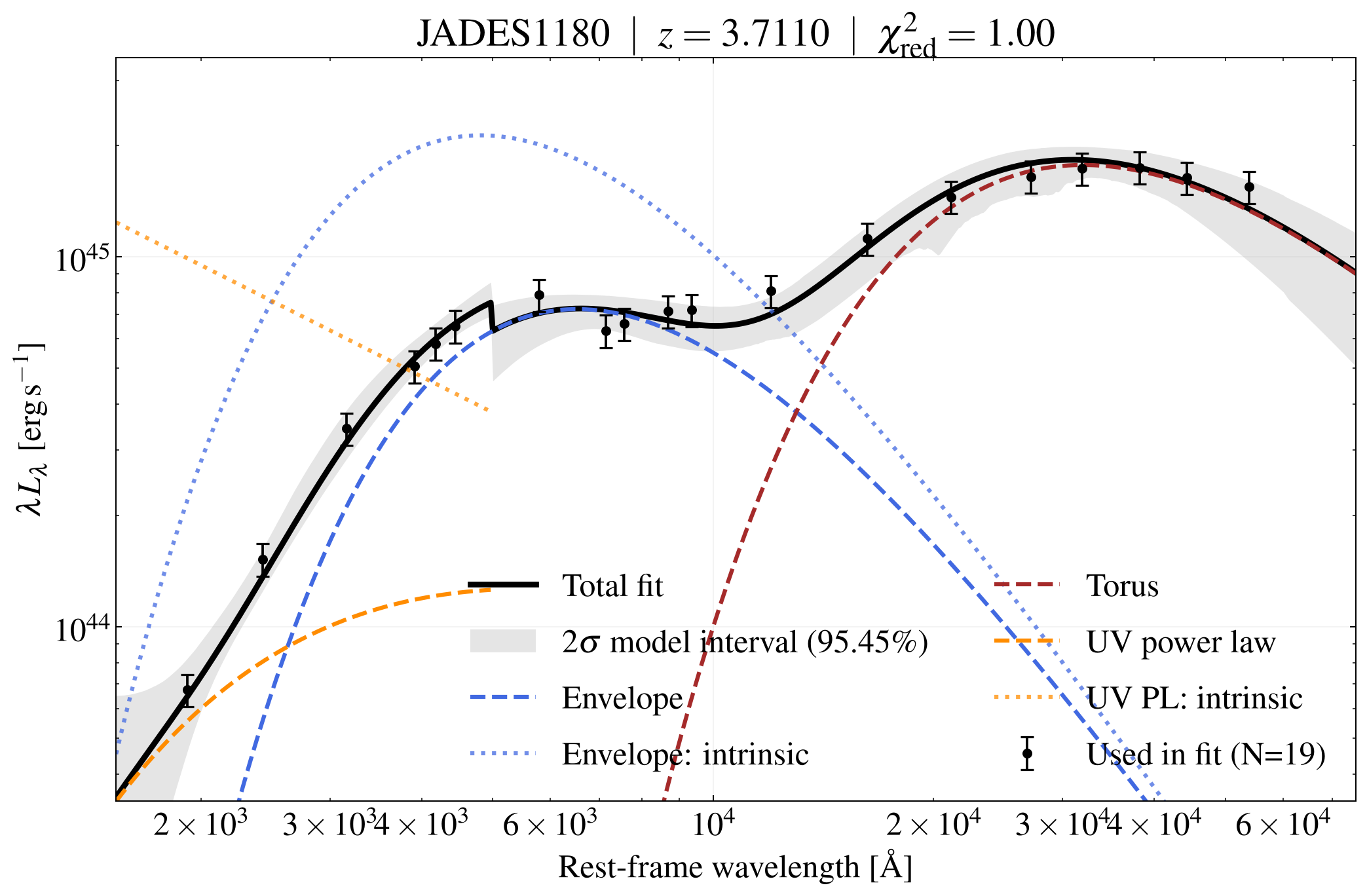}~
\includegraphics[width=0.5\textwidth]{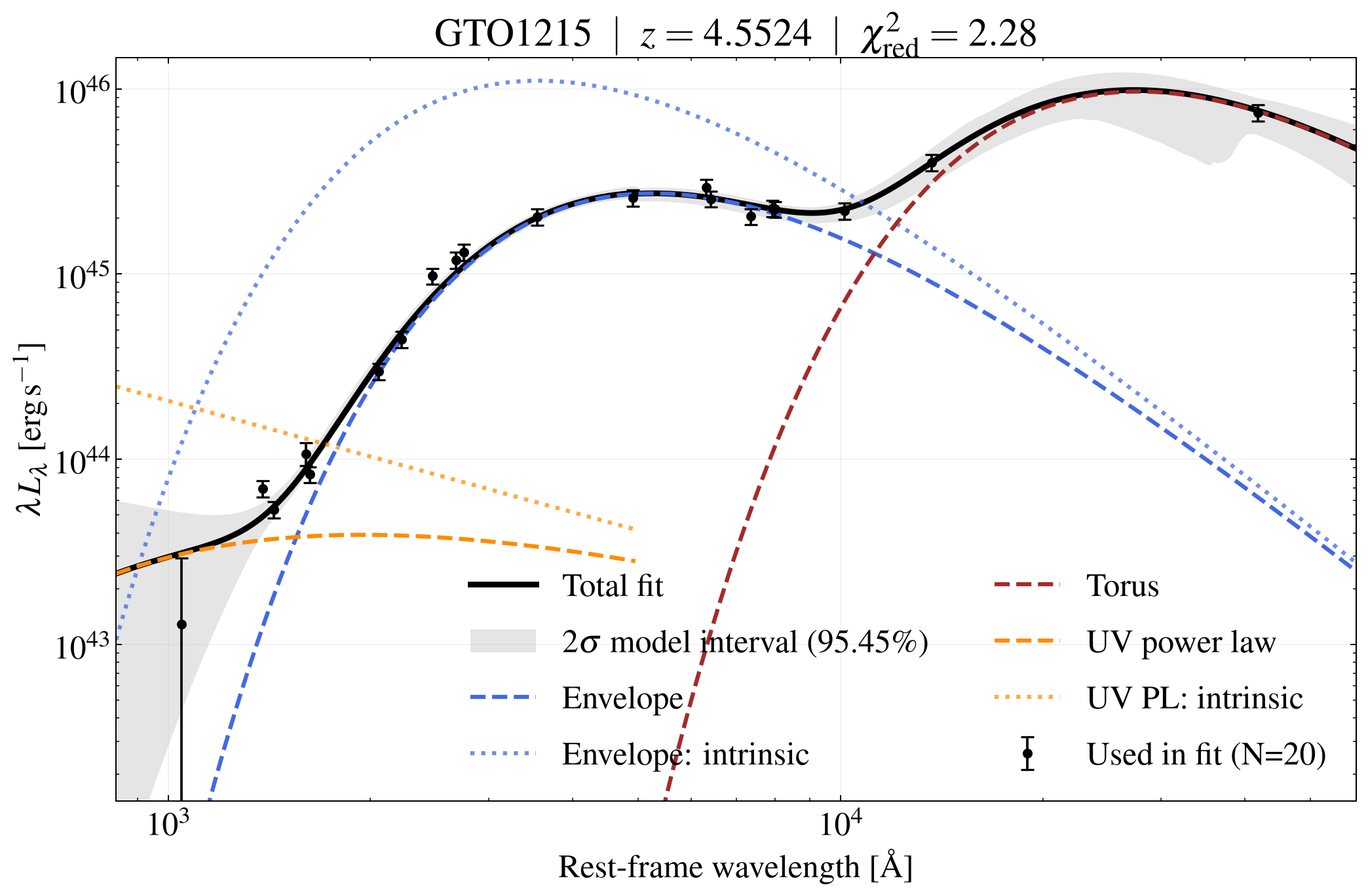}
\caption{
\textit{(Top)}
JWST/NIRSpec spectra (gray shaded region) of the two BBQs in this study, overlaid with the total model spectra (orange solid line), made with the combination of two blackbody components (blue dashed line for the envelope one and the brown dashed one for the AGN heated dust component) and \feii template (purple dashed lines).
\textit{(Bottom)}
Multi-wavelength SED fitting for UV-to-mid-IR photometry with two blackbody components representing the SMBH envelope (blue) and AGN dust torus (brown), assuming $\av=1.2$ to account for the energy balance between the two components, plus a UV power-law (PL) slope, up to $\lambda_{\rm rest}=5000$~\AA~with fixed $\tnr{\beta}{\lambda}=-2$ and $\av$ determined by the SED fitting.
}
\label{fig:SED}
\end{center}
\end{figure*}

We classified BBQs using the \(\beta_{\mathrm{UV}}\)–\(\beta_{\mathrm{opt}}\) plane
as shown in Figure~\ref{fig:betabeta}. 
This figure displays a cluster of points near the blue-blue location with $\beta_\mathrm{UV}\sim -2$ and $\beta_\mathrm{opt}\sim -2$, and two main sequences \citep{boo26}: 
one heading from the blue-blue location to the blue-red location where the LRDs are located ($\beta_\mathrm{UV}\sim -1.5$ and $\beta_\mathrm{opt}\sim 1$) along the vertical sequence, and another from the blue-blue location to the red-red location along the diagonal ($\beta_\mathrm{UV}\sim 2$ and $\beta_\mathrm{opt}\sim 0$).
In the red-red location, we then identified two sources with extremely red UV continua of $\beta_\mathrm{UV}>2.0$, but similar blue optical continua with non-LRDs of $\beta_\mathrm{opt}<-0.75$. These sources clearly deviate from the non-LRD and LRD sequences in the \(\beta_{\mathrm{UV}}\)–\(\beta_{\mathrm{opt}}\) plane. The selection of these two boundaries for the UV and optical slopes is arbitrary, but based on visual inspection conducted for the PRISM spectra of these two sources, including sources within their vicinity in the $\beta_\mathrm{UV}$--$\beta_{\mathrm{opt}}$ plane (sources with $\beta_\mathrm{UV}>2.0$ and $\beta_\mathrm{opt}<0.0$), they exhibited distinct spectra compared to the other sources. The selected sources in Figure~\ref{fig:betabeta} not only show characteristic $\Lambda$-shaped rest-frame UV-optical spectra, closely resembling a blackbody-like spectrum, but also clearly show broad \mgtwo \ and H$\alpha$ emission lines.
Therefore, we selected these two sources as a target of JWST BBQs in this study.
The first is \targetone\ at $z=3.711$ and the other is \targettwo\ at $z=4.552$.

\section{Results}\label{sec:results}

\subsection{JWST/NIRSpec Spectrum}\label{sec_sub:spectrum}
The spectra and spectral properties of the two sources are already reported in 
\cite{ma26} for \targetone\ and \cite{li26} for \targettwo, respectively. 
Thus we here summarize their fitting results (steep extinction curve assumption), our spectral fitting (blackbody spectral fitting) and key emission line properties used in this paper. 

Figure~\ref{fig:dustQSO} and the top panels of Figure~\ref{fig:SED} show the obtained JWST/NIRspec spectra (gray/black lines for Figure~\ref{fig:dustQSO} / Figure~\ref{fig:SED}).
The overall spectral features resembles the BBQSORS reported in \cite{zho26},
i) steep rising spectra at $\lambda_\mathrm{rest}<3000$~\AA, which is the natural outcome from the slope selection of $\beta_\mathrm{UV}>2.0$ shown in Figure~\ref{fig:betabeta},
ii) prominent broad \mgtwo~line with ${\rm FWHM}\gtrsim6000~\kms$, 
associated with narrow \mgtwodou doublet absorption at least for \targetone\ \citep{ma26}, iii)
$\Lambda$-shape continuum with a peak around $\lambda\approx3000$--$4000$~\AA, and iv) strong \feii pseudo continuum around $\sim2800$~\AA.

We first examine whether the observed spectra can be explained by dust extinction of an intrinsically blue quasar continuum. 
We normalize all spectra at $\lambda_{\rm rest}=4000$~\AA\ and fit the spectra of the two BBQs using a blue-quasar template \citep{sel16} attenuated by a Small Magellanic Cloud (SMC)-like dust extinction curve \citep{Gordon2024_dust}. 
We constrain $A_{\rm V}$ by requiring the reddened quasar template to reproduce the BBQ spectra at $\lambda_{\rm rest}>4000$~\AA, obtaining $\av=1.3$~mag for \targetone\ and $\av=0.6$~mag for \targettwo. Figure~\ref{fig:dustQSO} compares the spectra of the two BBQs with those of the blue-quasar template and the corresponding dust-reddened quasar models. 
The comparison shows that an SMC-like dust extinction curve cannot reproduce the steep drops of the BBQ spectra towards shorter wavelengths at $\lambda_{\rm rest}<4000$~\AA.
As discussed in \cite{li26}, this discrepancy can be solved only when we assume the extremely steep dust extinction curve with grain distribution of $q=3.8$ with $dn/da \propto a^{-q}$, which implies a grain mass distribution heavily skewed towards very small size at $a\sim0.01$~$\mu$m to produce a higher UV opacity.
We then fit the spectra with a reddened quasar template, adopting the steep extinction-curve parametrization given by Eq.~(1) of \cite{li26}. 
We allow the extinction $A_{\rm V}$, the extinction-curve parameters $c_1$, $c_2$, and $c_3$, and an overall normalization factor to vary simultaneously in the MCMC fitting. 
The best-fitting models yield $\av=1.0$~mag for \targetone\ and $\av=0.68$~mag for \targettwo.
In Figure~\ref{fig:dustQSO}, we compare the observed spectra with the corresponding dust-reddened quasar models and find that the steep extinction curves can reproduce the unusually red continua of both sources well.
We will discuss this point in Section~\ref{sec:BBQorigins}.

We then fit the spectra using a combination of two blackbody components representing emission from the gas envelope and AGN-heated dust, an \feii\, pseudo-continuum based on the I~Zw~1 template \citep{vest2001_feii,veron2004_feii,tsuzuki2006_feii}, and prominent emission lines using $\sanfit$ \citep{s3fit}. 
We broadly follow the fitting procedure of \cite{zho26}, who first reported a BBQ with significant radio detection, and refer the reader to that work for further details.
The allowed temperature ranges for the gas-envelope and dust blackbody components are $T_\mathrm{env,spec}=6000$--15,000~K and $T_\mathrm{dust,spec}=500$--1500~K, respectively.
The narrow-line components are modeled with values $\rm FWHM$ ranging from 50 to 1000~$\kms$, while the broad-line components are allowed to span 1000--10,000~$\kms$. 
A power-law extinction curve, defined as $A_{\lambda}\propto\lambda^{-1}$ and broadly consistent with an SMC-like extinction curve \citep{zho26}, is applied to the gas-envelope blackbody component with a fixed $A_{\rm V}=1.2$, consistent with the value adopted in the photometric SED fitting (see Section~\ref{sec: photo SED fitting}).
The top panels of Figure~\ref{fig:SED} present the fitting results (orange line) and both sources show a hotter envelope temperature of $T\sim10^4$~K, a roughly factor of two higher than that of typical LRDs, which again follows the result obtained in \citet{zho26} for BBQSORS at $z\sim1.7$. The full fitting result is summarized in Table~\ref{tab:properties}.

\subsection{UV to Mid-IR Photometry and SED fitting}
\label{sec: photo SED fitting}
We compiled multi-wavelength photometry from UV to mm (up to 1.24 mm), as well as X-ray bands, which are summarized in Table~\ref{tab:photometry}.
We perform an MCMC fitting for the obtained multi-wavelength SED covering from the UV to mid-infrared (MIR) wavelength range
up to $\lambda_\mathrm{rest}<10$~$\mu$m.
We first fit the SED using two modified blackbody (MBB) components representing the gas envelope and AGN-heated dust, respectively, adopting the same temperature ranges as those used in the spectral fitting.
Following \citet{deg25}, the MBB component is defined as
\[
F_\nu(\lambda)=N_{\rm MBB}B_\nu(T)
\left(\frac{\lambda_0}{\lambda}\right)^{\beta_{\rm MBB}},
\]
where $B_\nu(T)$ is the Planck function, $N_\mathrm{MBB}$ is the normalization, $T$ is the temperature, and $\beta_\mathrm{MBB}$ is the emissivity index. We adopt a pivot wavelength of $\lambda_0=5500$~\AA, corresponding to $\nu_0=c/\lambda_0$. A standard blackbody is recovered for $\beta_\mathrm{MBB}=0$.
For the low-temperature component that represents the AGN-heated dust, with $T_\mathrm{dust}\sim10^3$~K, we allow $\beta_\mathrm{MBB}$ to vary independently with $-2<\beta_\mathrm{MBB}<2$.
For the the gas envelope component with $T_\mathrm{env}\sim10^4$~K, we fix $\beta_\mathrm{MBB}=0$ and additionally apply the same power-law dust-extinction curve as in the spectral fitting, with a fixed $\av=1.2$~mag, representative of the values inferred for typical LRDs \citep[e.g.,][]{ume26}.
Fixing both $\av$ and $\beta_{\rm MBB}$ for this component reduces the degeneracy between effective temperature, emissivity, and dust extinction.
To account for the UV excess at $\lambda_\mathrm{rest}\lesssim2500$~\AA, we also included a power-law component with a fixed slope of $\beta_\lambda=-2$, defined as $F_\lambda\propto\lambda^{\beta_\lambda}$, consistent with that commonly observed in LRDs \citep{koc25}. 
This UV component may arise from a physically distinct region, such as the host galaxy, and may therefore experience a different level of dust attenuation from the gas envelope. 
We therefore allowed its extinction $\av$ to vary independently in the fitting.

Figure~\ref{fig:SED} shows the obtained UV to MIR SED and the best fitting results are overlaid with black solid curve. The best fitting result is summarized in Table~\ref{tab:properties}.
Similarly to the spectral fitting result, the rest-frame UV to optical band can be
fitted with a single blackbody spectrum representing an optically thick gaseous component and the values of $T_\mathrm{env}=7613_{-963}^{+806}$~K for \targetone\
and $T_\mathrm{env}=10336_{-239}^{+226}$~K for \targettwo, both values are factor of 1.5--2 higher than
that of typical LRDs with $T_\mathrm{env}\approx5000$~K.

The rest-frame near- to mid-IR SED shows a heated hot dust bump, which is usually absent or very weak in LRDs.
The rest-frame near- to mid-IR SED can be naturally reproduced by a blackbody of $T_\mathrm{dust}=1545_{-84}^{+41}$~K for \targetone\ and $T_\mathrm{dust}=1411_{-171}^{+132}$~K for \targettwo, whose value is very close to the dust sublimation temperature of $T_\mathrm{dust,sub}\approx1500$~K \citep{bar87}.
If dust emission is the reprocessed emission by absorbing the envelope UV to optical emission,
the luminosity ($L_\mathrm{dust}$) should not exceed the value of the original energy source ($L_\mathrm{env}$).
Although the observed SEDs show that the dust emission peak (red dashed line) is higher than that of the assumed envelope one (blue dashed line), the dust extinction corrected envelope peak (blue dotted line) for both sources have a higher peak than that of the dust emission,
and both sources fulfill $L_\mathrm{env,corr} \gtrsim L_\mathrm{dust}$ as summarized in Table~\ref{tab:properties}.

\begingroup
\setlength{\tabcolsep}{2pt}

\begin{deluxetable}{lcc}
\tabletypesize{\footnotesize}
\tablewidth{0pt}
\tablecaption{Summary of the photometry
\label{tab:photometry}}
\tablehead{
\colhead{Band} &
\colhead{\targetone} &
\colhead{\targettwo}
}
\startdata
HSC $r$
& \nodata
& $(2.78\pm0.73)\times10^{-5}$\textsuperscript{[1]} \\
HSC $i$
& \nodata
& $(8.14\pm0.76)\times10^{-5}$\textsuperscript{[1]} \\
HSC $z$
& \nodata
& $(1.46\pm0.21)\times10^{-4}$\textsuperscript{[1]} \\
ACS F275W
& $2.69\times10^{-6}$\textsuperscript{[7]}
& \nodata \\
ACS F336W
& $6.83\times10^{-6}$\textsuperscript{[7]}
& \nodata \\
ACS F435W
& $3.96\times10^{-6}$\textsuperscript{[7]}
& $(4.45\pm8.08)\times10^{-6}$\textsuperscript{[2]} \\
ACS F606W
& $1.90\times10^{-5}$\textsuperscript{[7]}
& $(1.14\pm0.45)\times10^{-5}$\textsuperscript{[2]} \\
ACS F775W
& $5.81\times10^{-5}$\textsuperscript{[7]}
& \nodata \\
ACS F814W
& $7.57\times10^{-5}$\textsuperscript{[7]}
& $(6.52\pm0.48)\times10^{-5}$\textsuperscript{[2]} \\
WFC3 F105W
& $2.49\times10^{-4}$\textsuperscript{[7]}
& \nodata \\
WFC3 F125W
& $6.17\times10^{-4}$\textsuperscript{[7]}
& $(8.42\pm0.33)\times10^{-4}$\textsuperscript{[3]} \\
WFC3 F140W
& $9.39\times10^{-4}$\textsuperscript{[7]}
& $(1.04\pm0.05)\times10^{-3}$\textsuperscript{[4]} \\
WFC3 F160W
& $1.39\times10^{-3}$\textsuperscript{[7]}
& $(2.75\pm0.03)\times10^{-3}$\textsuperscript{[4]} \\
JWST F090W
& $8.70\times10^{-5}$\textsuperscript{[2]}
& $1.14\times10^{-4}$\textsuperscript{[2]} \\
JWST F115W
& $3.63\times10^{-4}$\textsuperscript{[2]}
& $5.22\times10^{-4}$\textsuperscript{[2]} \\
JWST F150W
& $3.54\times10^{-4}$\textsuperscript{[2]}
& $2.70\times10^{-3}$\textsuperscript{[2]} \\
JWST F200W
& $2.67\times10^{-3}$\textsuperscript{[2]}
& $6.14\times10^{-3}$\textsuperscript{[2]} \\
JWST F277W
& $5.16\times10^{-3}$\textsuperscript{[2]}
& $1.07\times10^{-2}$\textsuperscript{[2]} \\
JWST F356W
& $5.64\times10^{-3}$\textsuperscript{[2]}
& $1.39\times10^{-2}$\textsuperscript{[2]} \\
JWST F410M
& $7.18\times10^{-3}$\textsuperscript{[2]}
& $1.28\times10^{-2}$\textsuperscript{[2]} \\
JWST F444W
& $7.76\times10^{-3}$\textsuperscript{[2]}
& $1.52\times10^{-2}$\textsuperscript{[2]} \\
\textit{Spitzer} $3.6\,\mu\mathrm{m}$
& $7.30\times10^{-3}$\textsuperscript{[5]}
& $1.58\times10^{-2}$\textsuperscript{[5]} \\
\textit{Spitzer} $4.5\,\mu\mathrm{m}$
& $9.53\times10^{-3}$\textsuperscript{[5]}
& $1.52\times10^{-2}$\textsuperscript{[5]} \\
\textit{Spitzer} $5.8\,\mu\mathrm{m}$
& $1.48\times10^{-2}$\textsuperscript{[5]}
& $1.89\times10^{-2}$\textsuperscript{[5]} \\
\textit{Spitzer} $8.0\,\mu\mathrm{m}$
& $3.07\times10^{-2}$\textsuperscript{[5]}
& $4.65\times10^{-2}$\textsuperscript{[5]} \\
\textit{Spitzer} $24\,\mu\mathrm{m}$
& $0.13$\textsuperscript{[6]}
& $2.64\times10^{-1}$\textsuperscript{[5]} \\
PACS $100\,\mu\mathrm{m}$
& \nodata
& $3.5\pm2.54$\textsuperscript{[6]} \\
PACS $160\,\mu\mathrm{m}$
& \nodata
& $5.7\pm3.97$\textsuperscript{[6]} \\
SPIRE $250\,\mu\mathrm{m}$
& $<17.4$\textsuperscript{[6]}
& $11.9\pm2.4$\textsuperscript{[6]} \\
SPIRE $350\,\mu\mathrm{m}$
& $<21.6$\textsuperscript{[6]}
& $16.1\pm2.7$\textsuperscript{[6]} \\
SPIRE $500\,\mu\mathrm{m}$
& $<2.2$\textsuperscript{[6]}
& $6.8\pm4.4$\textsuperscript{[6]} \\
ALMA $1.24\,\mathrm{mm}$
& $0.16$\textsuperscript{[8]}
& \nodata \\
VLA 1.5~GHz
& \nodata
& $<5.03\times10^{-2}$\textsuperscript{[9,11]} \\
VLA 3~GHz
& $(1.98\pm0.10)\times10^{-3}$\textsuperscript{[10]}
& \nodata \\
VLA 6~GHz
& $<0.93\times10^{-3}$\textsuperscript{[10]}
& \nodata \\
\hline
Chandra 0.5--2~keV
& $(8.52 \pm 0.31)\times10^{-16}$\textsuperscript{[11]}
& $(1.70 \pm 0.21)\times10^{-15}$\textsuperscript{[11]}\\
\enddata
\tablecomments{
Flux densities are in mJy, except for the X-ray fluxes.
Chandra fluxes are measured over $0.5$--$2\,\mathrm{keV}$
and are in $\mathrm{erg\,s^{-1}\,cm^{-2}}$.
Bracketed superscripts identify references.
HSC is on Subaru; ACS and WFC3 are on HST;
PACS and SPIRE are on Herschel.
Radio flux densities assume unresolved emission.
The VLA 1.5 and 6~GHz upper limits are three times the local RMS.
The 1.5~GHz limit is derived in this work from the public
image and RMS map of \citet{hey20}.
References:
[1] \citet{straatman_fourstar};
[2] \citet{astrodeep};
[3] \citet{galametz2013};
[4] \citet{skelton2014};
[5] \citet{spitzer_seip};
[6] \citet{oliver2012_herschel};
[7] \citet{whitaker2019_hlf};
[8] \citet{fujimoto2017_alma};
[9] \citet{hey20};
[10] \citet{ma26};
[11] This work.
}
\end{deluxetable}
\endgroup

\begingroup
\setlength{\tabcolsep}{2pt}
\begin{deluxetable}{lcc}
\tabletypesize{\footnotesize}
\tablewidth{\columnwidth}
\tablecaption{Summary of fitting results and physical properties
\label{tab:properties}}
\tablehead{
\colhead{Property} &
\colhead{\targetone} &
\colhead{\targettwo}
}
\startdata
$z$ & $3.711$ & $4.552$ \\
RA [deg] & $53.1584709$ & $34.33731548$ \\
Dec [deg] & $-27.7740461$ & $-5.1436736$ \\
\hline
\multicolumn{3}{c}{Spectroscopy} \\
\hline
$T_{\rm env,spec}$ [K] & $8214 \pm 73$ & $11120 \pm 767$ \\
$L_{\rm env,spec}$ [erg~s$^{-1}$] & $(2.97 \pm 0.05)\times10^{45}$ & $(1.36 \pm 0.08)\times10^{46}$ \\
FWHM (BLR) [km~s$^{-1}$] & $5858 \pm 139$ & $6119 \pm 507$ \\
$F({\rm C\,III]})$ & $6.89 \pm 2.23$ & $5.88 \pm 8.89$ \\
$F({\rm Mg\,II})$ & $7.54 \pm 3.00$ & $31.2 \pm 8.3$ \\
$F({\rm H}\beta,{\rm broad})$ & $46.1 \pm 2.6$ & $122 \pm 35$ \\
$F({\rm H}\alpha,{\rm broad})$ & $260 \pm 3$ & $441 \pm 83$ \\
$M_{\rm BH,Mg\,II}$ [$M_{\odot}$] & $1.60^{+4.08}_{-1.15}\times10^{8}$ & $4.45^{+11.5}_{-3.21}\times10^{8}$ \\
$\lambda_{\rm Edd,Mg\,II}$
& $0.260^{+0.663}_{-0.187}$
& $0.507^{+1.349}_{-0.369}$ \\
$M_{\rm BH,H\alpha}$ [$M_{\odot}$]
& $(8.96 \pm 0.46)\times10^{8}$
& $(1.50 \pm 0.35)\times10^{9}$ \\
$\lambda_{\rm Edd,H\alpha}$
& $0.046 \pm 0.002$
& $0.130 \pm 0.024$ \\
\hline
\multicolumn{3}{c}{Photometry} \\
\hline
$T_{\rm env,ph}$ [K]
& $7613^{+806}_{-963}$
& $10336^{+226}_{-239}$ \\
$L_{\rm env,ph}$ [erg~s$^{-1}$]
& $2.89^{+0.40}_{-0.39}\times10^{45}$
& $1.50^{+0.05}_{-0.06}\times10^{46}$ \\
$T_{\rm torus}$ [K]
& $1545^{+41}_{-84}$
& $1411^{+132}_{-171}$ \\
$L_{\rm torus}$ [erg~s$^{-1}$]
& $2.93^{+0.17}_{-0.17}\times10^{45}$
& $1.37^{+0.12}_{-0.12}\times10^{46}$ \\
\hline
\multicolumn{3}{c}{X-ray Properties} \\
\hline
$L_{2-10\,{\rm keV,int}}$ [erg~s$^{-1}$]
& $(1.30 \pm 0.06)\times10^{44}$
& $4.5^{+1.2}_{-1.0}\times10^{44}$ \\
$N_{\rm H}$ [cm$^{-2}$]
& $<2.4\times10^{22}$
& $<1.5\times10^{23}$ \\
\enddata
\tablecomments{
Line fluxes are in units of
$10^{-18}\ {\rm erg\,s^{-1}\,cm^{-2}}$.
}
\end{deluxetable}
\endgroup

\subsection{Far-IR Observations}\label{sec:FIR}

Table~\ref{tab:photometry} summarizes that both sources have detection either in Herschel/PACS+SPIRE or ALMA~1.24~mm, indicating that their rest-frame far-IR emissions are detected. Since their fluxes are significantly brighter than \textit{Spitzer}/IRAC or MIPS band, their emission cannot be described by hot dust emission shown in the bottom panels of Figure~\ref{fig:SED}. This indicates that their emission would likely originate from the host galaxy emission, which is again usually non-detection for LRDs. We will discuss those points later in Section~\ref{sec:comparison_LRDBBQ}.

\subsection{Radio Observations}\label{sec:radio}

\targetone\ is in the GOODS-South field and 
we adopt the VLA peak flux densities reported by \cite{ma26}, with
$f_\mathrm{3GHz} = 1.98\pm0.10~\mu$Jy and $f_\mathrm{6GHz}<0.93~\mu$Jy at 3 and 6~GHz, respectively,
and summarized in Table~\ref{tab:photometry}. 
The upper-limit provides the expected radio slope for \targetone\ is relatively steep, with $\alpha<-1.1$ with $f_\nu \propto \nu^\alpha$.

\targettwo\ is located in the UDS field, which is within the VLA 1--2~GHz XMM-LSS survey \citep{hey20}.
From the public continuum mosaic data at $\sim1.5$~GHz and its associated RMS map, we measure the intensity of the target-position pixel and the upper-limit value of 3$\sigma$ is $f_\mathrm{1.4GHz}<50.3~\mu$Jy.

\begin{figure*}[tp!]
\begin{center}
\includegraphics[width=0.16\textwidth]{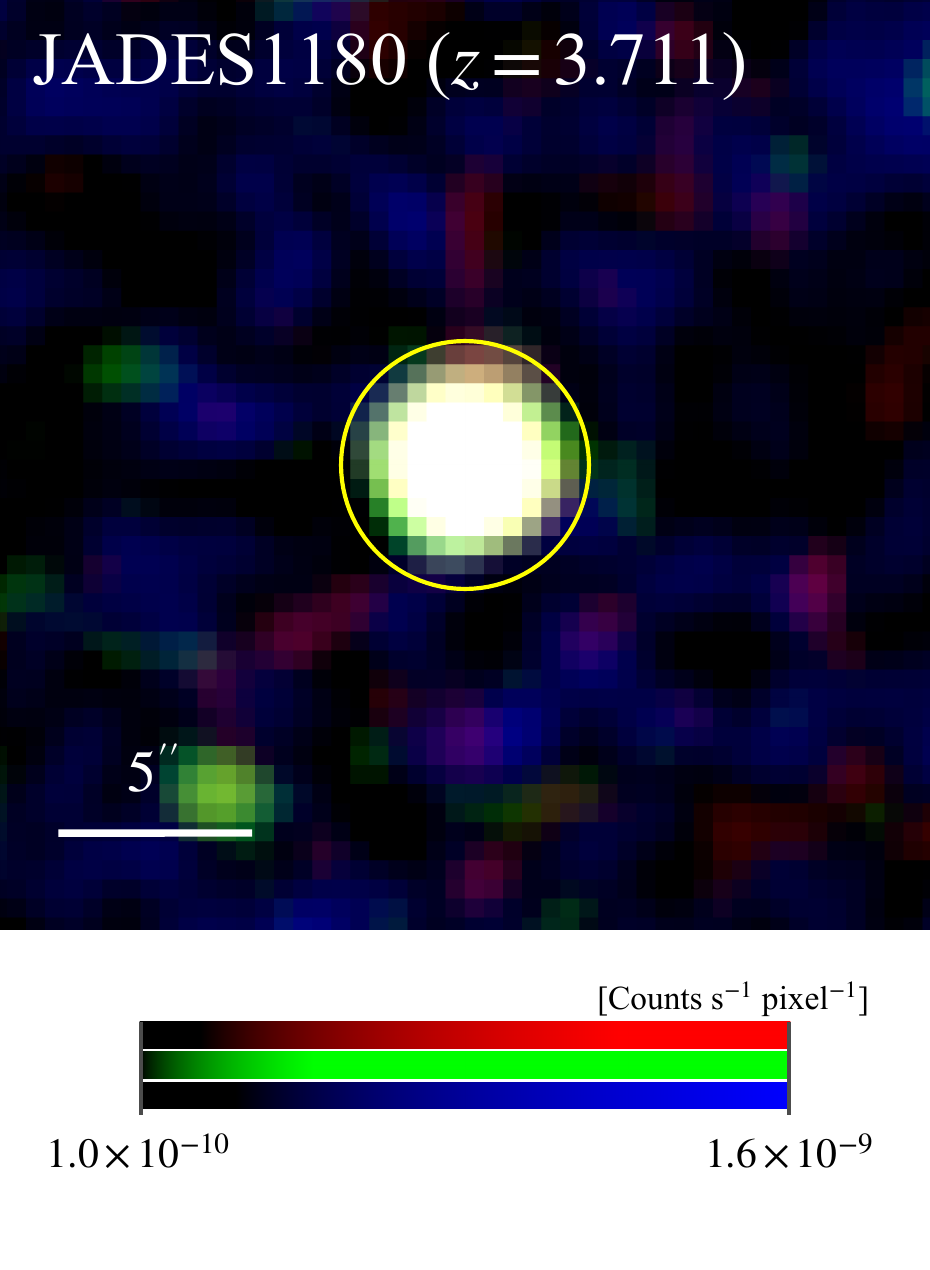}~
\includegraphics[width=0.32\textwidth]{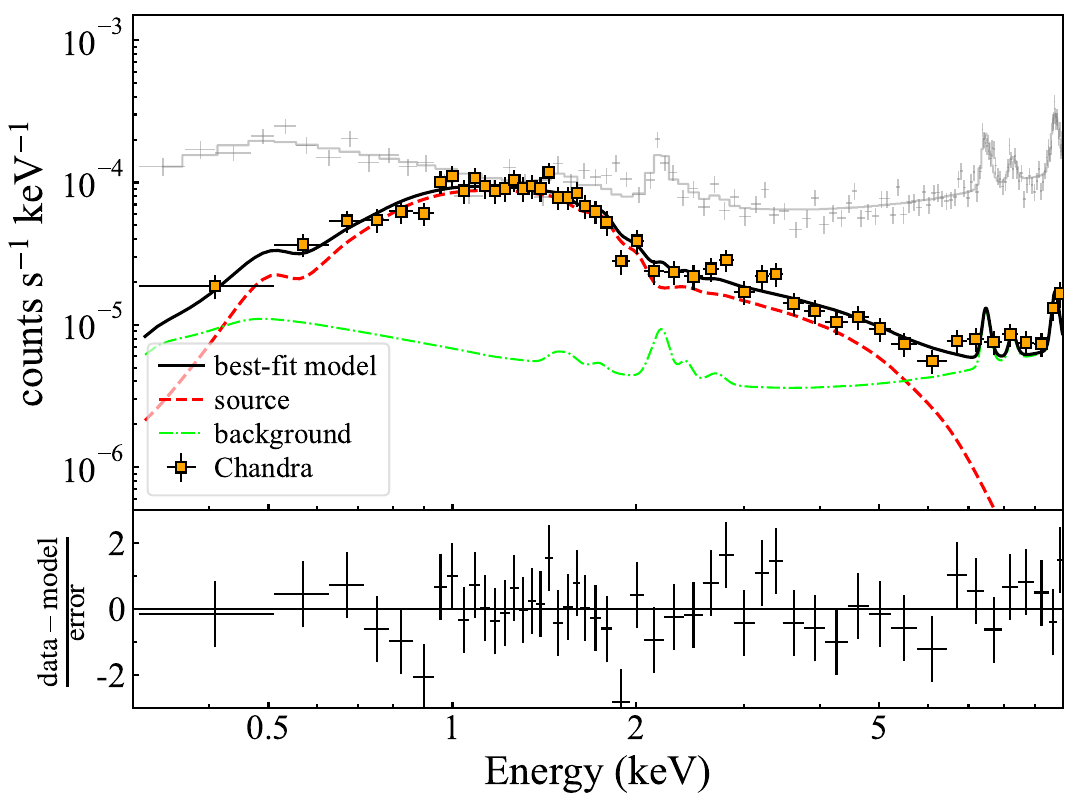}~
\includegraphics[width=0.16\textwidth]{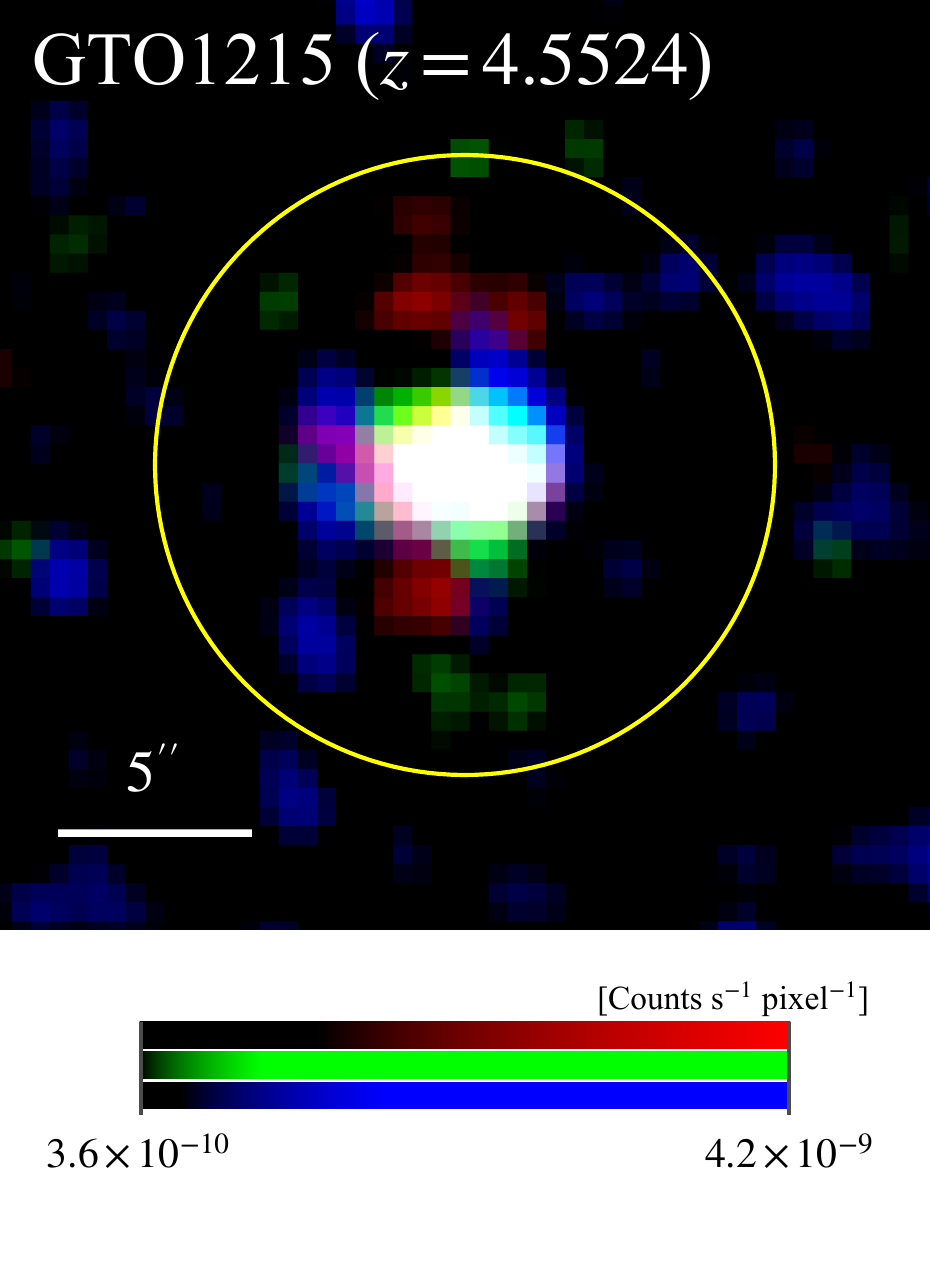}~
\includegraphics[width=0.32\textwidth]{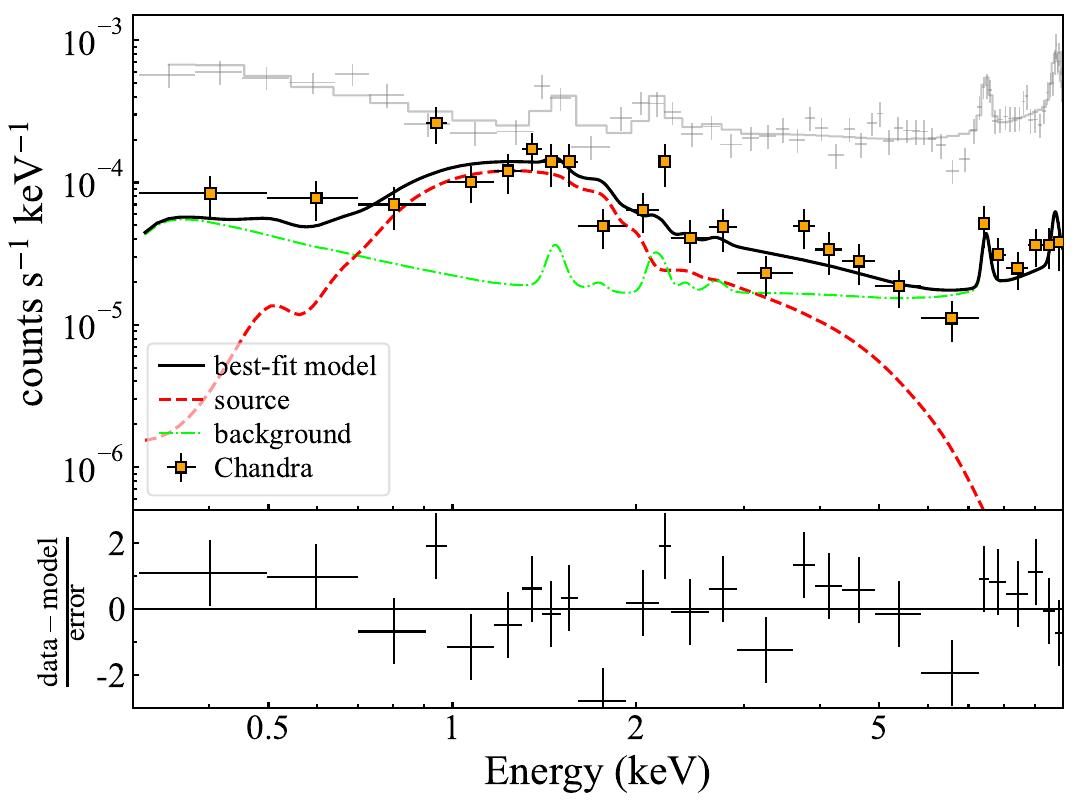}
\caption{
X-ray detections of two BBQs with Chandra.
The left panels show three-color images in the 0.5--1.2 (red), 1.2--2 (green), and 2--7 keV (blue) bands, with yellow circles marking the typical source radii (3.2$''$ for \targetone\ and 8$''$ for \targettwo). The right panels show the 0.3--10 keV spectra (rest-frame 1.4--47 keV and 1.7--56 keV, respectively), with the best-fitting model (black), source model (red), and area-scaled background model (green). Background data and model are shown in gray.}
\label{fig:Xray}
\end{center}
\end{figure*}

\subsection{X-ray Analysis}\label{sec:X-ray}
We analyzed deep Chandra observations of the two objects.
\targetone\ was observed in the 7 Ms Chandra Deep Field-South \citep[CDF-S;][]{luo17} between 1999 and 2016, with 111 pointings, while \targettwo\ was observed in the X-UDS field between 2010 and 2015, with 26 pointings \citep{koc18}. 
Because the nominal focal-plane temperature was ${-}110$~degrees Celsius for observations taken before 2000, compared to ${-}120$~degrees Celsius for later observations, we excluded the two observations from 1999. 
All data were reduced with CIAO v4.18 and CALDB v4.12.4 using \texttt{chandra\_repro}, and periods of enhanced background were removed using an iterative 3$\sigma$-clipping approach \citep[e.g.,][]{luo17}.

We performed astrometric corrections for the observations. X-ray sources were detected in the 0.5--7.0 keV band with \texttt{wavdetect}, adopting a false-positive probability threshold of $10^{-6}$ and the $\sqrt{2}$ sequence of wavelet scales from 1 to 8 pixels. 
Point-spread function maps were generated with \texttt{mkpsfmap} assuming an encircled count fraction (ECF) of 0.393, a power-law spectrum with $\Gamma = 1.4$, and Galactic hydrogen column densities of $7.18\times10^{19}$ cm$^{-2}$ and $2.18\times10^{20}$ for the CDF-S and X-UDS fields, respectively
\citep{wil13}.
Using \texttt{reproject\_aspect}, we matched the positions of the X-ray sources with those of the Chandra Source Catalog (CSC; Evans et al. 2024) within a 3$''$ matching radius, adopting a residual limit of 0.6$''$.
We merged the event files for each field using \texttt{merge\_obs}. 
The resulting angular offsets relative to the CSC positions are ${<}0.1''$, while the absolute astrometric accuracy is ${\sim}0.3''$ at the 95\% confidence level \citep{eva24}.

Source events were extracted from a circular region centered on the source position, with a radius corresponding to a 95\% ECF (typically 3.2$''$ and 8$''$ for \targetone\ and \targettwo, respectively).
Background events were extracted from an annular region with an inner radius 1.5 times the source radius and an outer radius 40 pixels larger than the inner radius, excluding nearby X-ray sources with a radius of 14$''$ \citep[e.g.,][]{koc18}.
To maximize the detection significance, we selected observations with angles off the axis of ${<}10'$ and excluded low-exposure observations that yielded negligible counts at the target positions (ObsIDs 16186, 16187, 17634, and 18709).
This left 98 and 13 observations, with net exposures of 6.78 Ms and 629 ks and 0.5--2.0~keV fluxes of $8.5\times10^{-16}$ and $1.7\times10^{-15}$ erg~cm$^{-2}$~s$^{-1}$ for \targetone\ and \targettwo, respectively.

Using the $C$-statistic, we simultaneously fit the source and background spectra in the 0.3--10.0~keV (rest-frame $\sim$2--50~keV) band.
We employed a background model consisting of Gaussian emission lines, a power law, and an exponentially decaying component with Gaussian absorption, with the normalizations and the power-law photon index left free and the other parameters fixed to those of \cite{uem25}, who analyzed ACIS-I data taken in VFAINT mode.
The \texttt{gabs*powerlaw} model was also included to account for the background in the ACIS-S or FAINT-mode observations, with its parameters left free \citep[e.g.,][]{bar14,suz21}.
The spectra are well described by a Galactic-absorbed power law (\texttt{TBabs*zpowerlaw}), yielding best-fit photon indices of $\Gamma=1.74\pm0.07$ and $2.21^{+0.34}_{-0.28}$ for \targetone\ and \targettwo, respectively.\footnote{The photon index of \targetone\ is consistent with that reported by \cite{luo17}, $\Gamma=1.71^{+0.08}_{-0.07}$.}
We find 3$\sigma$ upper limits on intrinsic absorption of $N_{\rm H}<2.4\times10^{22}$ and $<1.5\times10^{23}$~cm$^{-2}$ using \texttt{zTBabs}. 
The corresponding intrinsic 2--10~keV luminosities are $1.30\pm0.06\times10^{44}$ and $4.5^{+1.2}_{-1.0}\times10^{44}$~erg~s$^{-1}$, respectively.
The merged 0.5--7.0~keV data yield detection significances of $27.7\sigma$ and $8.1\sigma$, respectively, and the corresponding images and spectra are shown in Figure~\ref{fig:Xray}.
These results indicate that both sources are consistent with the non Compton-thick ($\NHunit<10^{24}$) sources, whose extremely gas obscurations are expected for LRDs \citep{ana24}.

\begin{figure*}
    \centering
    \includegraphics[width=0.45\textwidth]{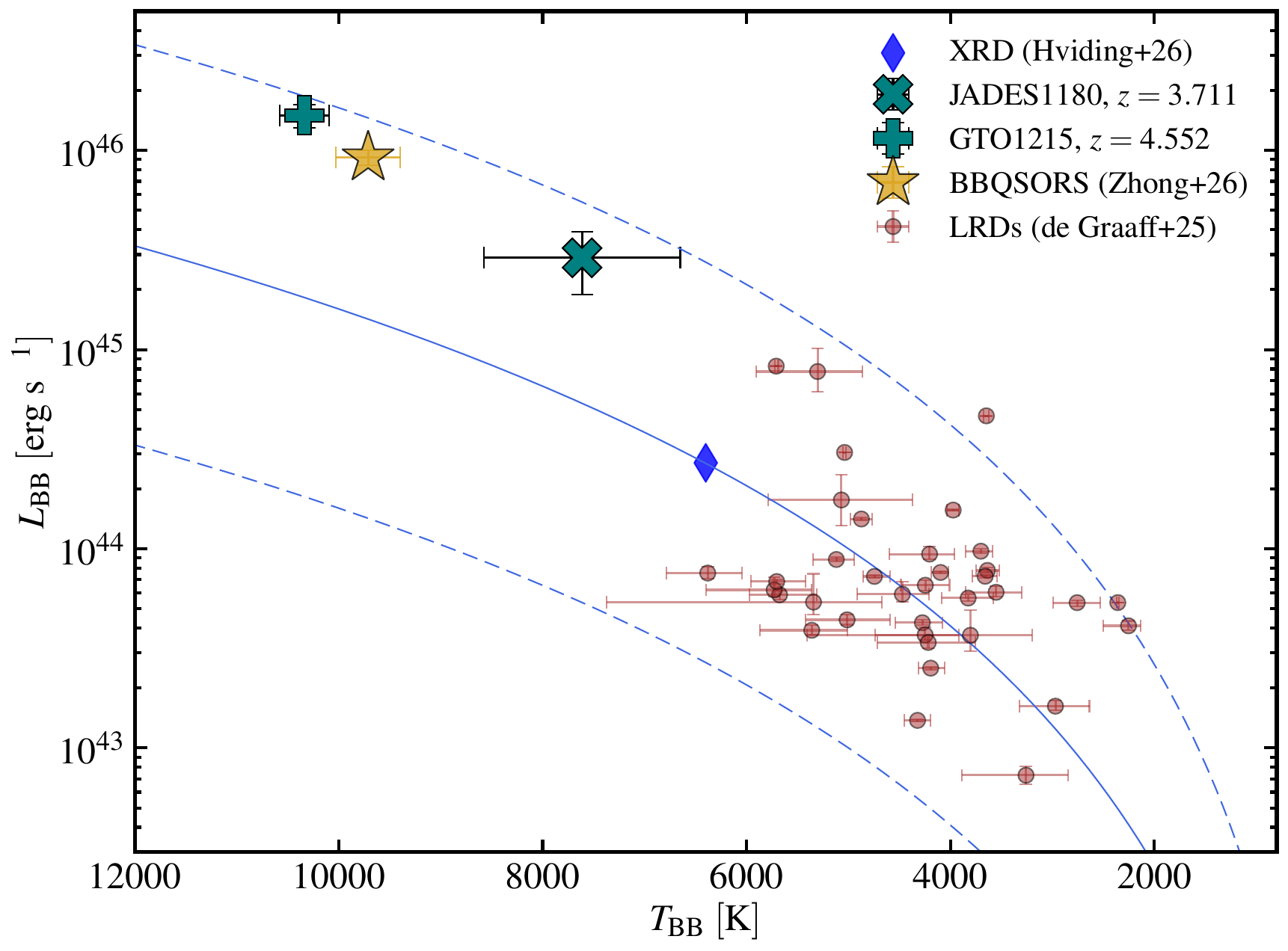}~
    \includegraphics[width=0.45\textwidth]{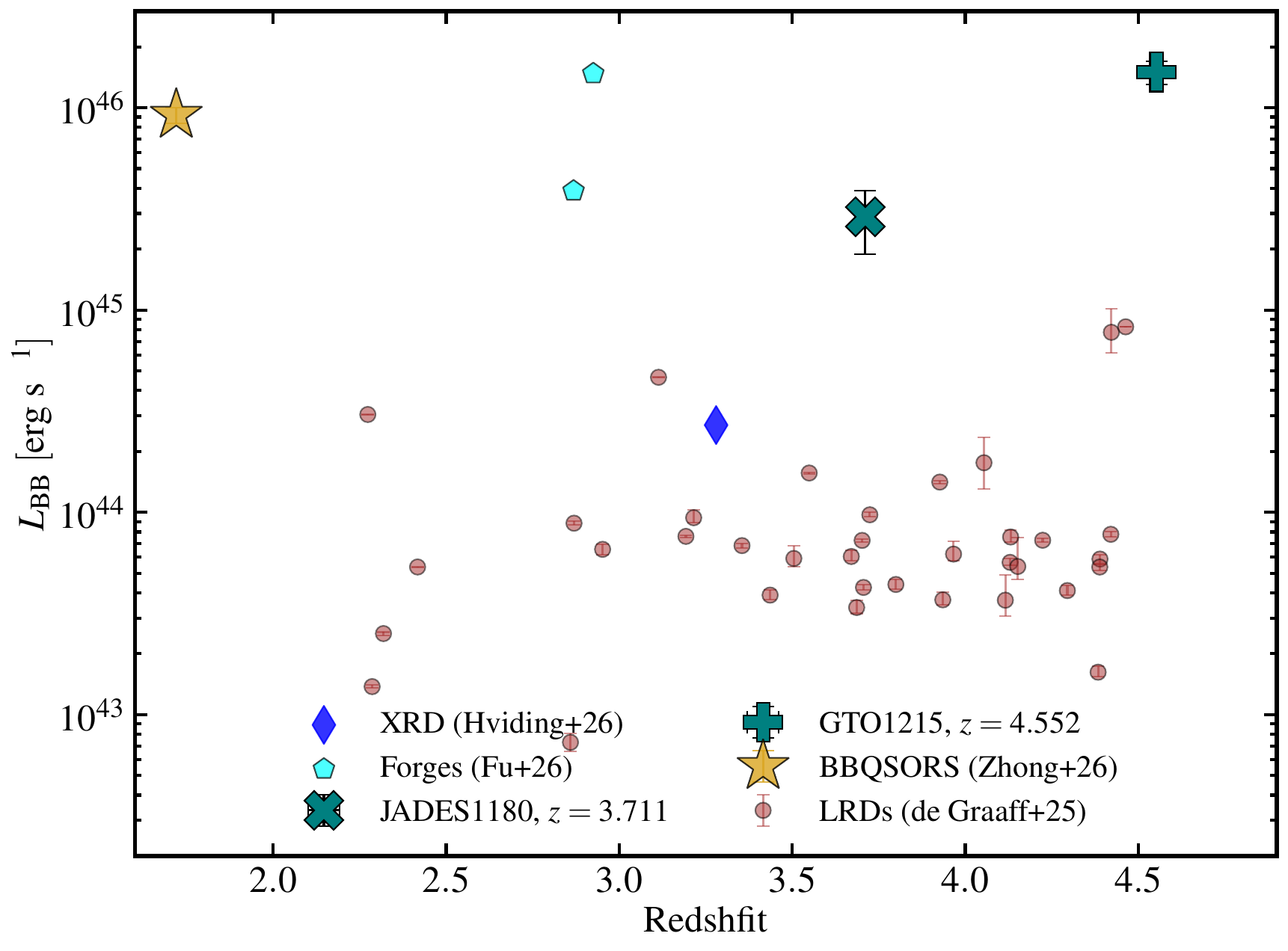}\\
    \caption{
    Comparison of \targetone\ and \targettwo\ with LRDs at $z<4.5$ from \citet{deg25}, and transitional LRD candidates, the BBQSORS \citep{zho26}, XRD \citep{hvi25} and Forges \citep{fu25}, in $L_\mathrm{BB}$--$T_\mathrm{BB}$, and $L_\mathrm{BB}$--$z$ planes.
    In the left panel, the solid line indicates the expected $\tnr{L}{BB}\text{--}\tnr{T}{BB}$ relation assuming a fiducial LRD-like photospheric radius \citep{ina26}, with dashed lines are offset by $\pm1$~dex.}
    \label{fig:LRDcomparison}
\end{figure*}

\begin{figure}
    \centering
    \includegraphics[width=0.95\columnwidth]{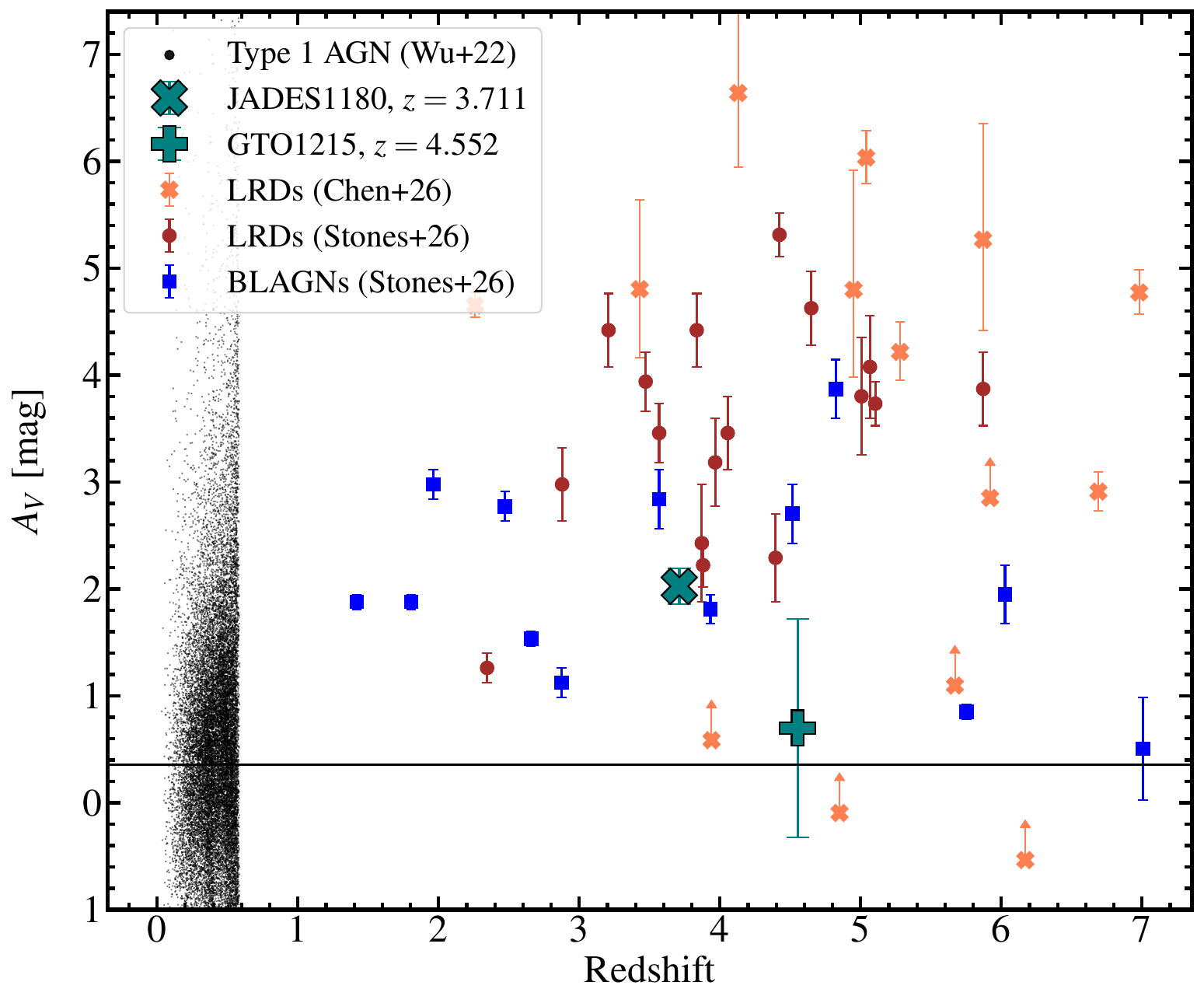}
    \caption{
    Extinction $A_{\rm V}$ estimated from observed Balmer decrement for two BBQ sources, LRDs \citep{stones26_lrd,chen26_abcd}, broad-line AGNs \citep{stones26_lrd}, and low-$z$ Type 1 AGNs \citep{wu22_sdssqso}.
    Estimates are based on the total integrated fluxes from each line for samples from \cite{stones26_lrd}, while the other sample use only the broad components.
    The horizontal solid line represents the median $A_{\rm V}$ of Type 1 AGNs.
    }
    \label{fig: av vs z}
\end{figure}

\section{Discussion}\label{sec:discussion}

\subsection{Comparison between BBQs and LRDs: implications for the origins of the UV-optical continuum}\label{sec:comparison_LRDBBQ}

The two BBQs share several properties with LRDs, including compact morphology (similar photospheric radius as shown below), broad Balmer emission lines, and a UV-to-optical continuum that can be described by blackbody-like emission from a dense gas envelope. 
On the other hand, BBQs also differ from LRDs in several respects: a $\Lambda$-shaped rather than a ``V''-shaped UV-optical continuum, a hotter blackbody temperature ($T_\mathrm{env}\sim10^4$~K), and the presence of host dust ($T_\mathrm{dust}\sim10^3$~K) and X-ray emission.
Instead, these distinctive features are shared with BBQSORS reported by \cite{zho26}, although \targettwo\ is not detected in the radio regime. 
This combination distinguishes BBQs from LRDs and motivates consideration of a possible evolutionary connection between the fully gas-enshrouded phase of LRDs and classical quasars.

One notable difference between BBQs and LRDs is their envelope temperature $T_\mathrm{env}$ contrast.
The left panel of Figure~\ref{fig:LRDcomparison} shows the location of BBQs and LRDs in the $L_\mathrm{env}$--$T_\mathrm{BB}$ plane.
Whereas LRDs cluster at $T\sim5000$~K and $L_\mathrm{env}<10^{45}$~erg~s$^{-1}$, the two BBQs have $T_\mathrm{env}\sim7500$--$10000$~K and $L_\mathrm{env}>10^{45}$~erg~s$^{-1}$, exceeding the cut-off luminosities of the LRDs and thus their BH masses would reach $M_\mathrm{BH}>10^7 \msun$ assuming Eddington limit \citep{ma25b}. 
On the other hand, their positions remain consistent with the sequence of constant photospheric radius illustrated in Figure~\ref{fig:LRDcomparison}.
For fiducial LRD values of $L_\mathrm{env} = 10^{44}\ \ergs$, and $T_\mathrm{BB}=5000$~K, their photospheric radius is $R_\mathrm{ph,LRD} = \sqrt{L_\mathrm{env}/4\pi\sigma_\mathrm{SB}T_\mathrm{env}^4} \sim 5 \times 10^{-3}$~pc \citep{ina26},
where $\sigma_\mathrm{SB}$ is the Stefan-Boltzmann constant. Since $R_{\rm ph} \propto (L_\mathrm{env})^{1/2}/(T_\mathrm{env})^{2}$, the $\sim$10 times higher $L_{\rm env}$ and $\sim$2 times higher $T_{\rm env}$ of the BBQs roughly compensate each other ($\sqrt{10}/2^{2} \approx 0.8$), giving $R_{\rm ph}$ of the same order as LRDs.
Thus, under the envelope interpretation, the envelope radii of BBQs remain comparable to those of LRDs.

Recently, a physical framework for the connection between LRD and BBQs is provided by \citet{wan26}, who model the coupled evolution of a growing BH and its gaseous envelope under declining gas supply, radiative cooling, and accretion-powered heating. 
\citet{wan26} showed that while the cool envelope with $T\sim5000$~K initially remains near the Hayashi track, as the gas supply becomes insufficient to replenish the envelope, the photosphere starts to heat, producing an intermediate phase with
$T_\mathrm{eff}\sim10^4$~K before the system evolves toward a Little Blue Dot (LBD) or blue quasar-like continuum. 
The temperatures inferred for our BBQs are compatible with this intermediate thermal state.

For the point of the envelope geometry or thickness, the X-ray detections for BBQs provide additional information on the obscuration of the gas envelope, with Compton-thin level with $N_\mathrm{H} \lesssim 10^{23}$~cm$^{-2}$ as discussed in Section~\ref{sec:X-ray}.
If the optical continuum originates predominantly in an optically thick gas envelope, such X-ray visibility suggests incomplete gas covering or a possible clumpy gas envelope structure,
with a  possible partial envelope dispersal from the fully covered gas envelope shown in many LRDs. In addition, both sources show hot dust emission with $T_\mathrm{dust}\approx1500$~K, indicating that nuclear UV radiation should reach the surrounding dust.
Within the gas envelope interpretation, the dust would be heated either by reprocessed envelope emission, or directly escaping accretion emission through the possible partially dispersed area, or both. Both sources currently shows $L_\mathrm{env}\gtrsim L_\mathrm{dust}$, thus the only the former component is sufficient from the point of view of energy budget.

The other notable point is their far-IR (FIR) detections.
As discussed in Section~\ref{sec:FIR}, the FIR-submm measurements exceed
the extrapolation of the hot-dust component as shown in Table~\ref{tab:photometry}, indicating a contribution from cooler dust, potentially associated with the host galaxies.
Since the host galaxy properties of LRDs are still poorly explored, BBQs can be a good laboratory for their host galaxy environment. 

Such a possible gas envelope dispersion or thinner gas envelope is also partially inferred from the line ratio of broad emission lines.
We estimate the dust extinction $A_{\rm V}$ from the observed Balmer decrements, which is defined as \citep{dom13_balmer}:
\begin{equation}
    A_{\rm V}=\frac{2.5}{\kappa({\rm H\beta})-\kappa({\rm H\alpha})}\log_{10}\left[\frac{({\rm H\alpha/H\beta)_{obs}}}{({\rm H\alpha/H\beta)_{int}}}\right],
\end{equation}
where $\kappa({\rm H\alpha})\approx1.19$ and $\kappa({\rm H\beta})\approx0.83$ are extinction coefficients utilizing an SMC-like extinction curve \citep{Gordon2024_dust}.
$({\rm H\alpha/H\beta)_{obs}}$ is the observed Balmer decrement, where only the broad components (see Table~\ref{tab:properties}) are considered, and  $({\rm H\alpha/H\beta)_{int}}=2.86$ is the intrinsic Balmer decrement assuming a Case B recombination.

Figure~\ref{fig: av vs z} compares the estimated $A_{\rm V}$ values of our two BBQ sources with those of LRDs \citep{stones26_lrd,chen26_abcd}, broad-line AGNs \citep{stones26_lrd}, and low-$z$ Type 1 AGNs \citep{wu22_sdssqso}.
With the assumption of case B, LRDs show extremely high $A_\mathrm{V}$ values reaching $A_\mathrm{V}\sim5$~mag, which deviates from the estimates from the gas envelope of $A_\mathrm{V}\approx1$~mag \citep[e.g.,][]{nik26}. This deviation can be obtained if the gas envelope condition is deviated from case B situation, reaching an extremely high density with $n_\mathrm{e}\gg10^6$~cm$^{-2}$ \citep{dra80}.
On the other hand, the two BBQ sources show relatively moderate $A_\mathrm{V}$ values with $A_\mathrm{V}\approx1-2$~mag, which is roughly consistent with the value expected from the LRD-like gas envelope. 
This naturally means that the envelope density is getting thinner for these two BBQs, possibly consistent with type-1 AGN level \citep{stones26_lrd}.

Beyond these comparisons of individual source properties, the redshift distribution of BBQs provides an opportunity to investigate their possible connection with the LRD population.
The right panel of Figure~\ref{fig:LRDcomparison} shows that the known BBQs occupy a unique region characterized by $L_\mathrm{env}>10^{45}$~erg~s$^{-1}$ over $1<z<5$.
Together with the previously reported BBQSORS at $z=1.7$, our discovery of two additional BBQs extends the population to $3<z<5$, now overlapping the redshift range of main LRD populations at $z=3$--$5$. These sources therefore provide an opportunity to investigate a possible transitional phase between LRDs and classical quasars at a similar cosmic epoch.

As BBQs could represent transitional states that LRDs can turn into, their relative abundance may allow us to calculate crude constraints on their duration. 
Although the JWST/NIRSpec sample is subject to the target selection biases, 
comparing the number of BBQs detected (2) with LRDs ($\sim$60) in \citet{boo26} and adopting the inferred timescale of LRDs $(t_{\rm LRD}\sim3\times10^7\,{\rm yrs})$ gives $t_{\rm BBQ}\approx t_{\rm LRDs}\frac{N_{\rm BBQ}}{N_{\rm LRD}}=1\,{\rm Myr}$, suggesting a very short-lived phase of envelope dispersal.

\subsection{Origins of $\Lambda$-shape in BBQs: Envelope scenario vs. Steep dust extinction}\label{sec:BBQorigins}

Although the properties of the two BBQs are compatible with the gaseous-envelope interpretation, blue quasar and dust extinction by a steep dust-extinction curve provides an alternative explanation for the observed UV--optical continua, as shown in Section~\ref{sec_sub:spectrum} and proposed by \cite{li26}.
They require an extremely steep dust extinction curve with $q=3.8$ with $dn/da \propto a^{-q}$ and its physical origins are not yet explored; therefore, the failure of an SMC-like extinction curve does not exclude the dust reddening scenario.
Actually, the X-ray detections, hot-dust emission, and additional FIR component can be described with a dust-reddened quasar embedded in a dusty host galaxy with such a steep extinction curve, and several galaxies are known to have such extreme end \citep[e.g., ][]{jia13,zaf15}. 
The principal distinction between the scenarios is whether the UV--optical continuum is dominated by thermalized envelope emission or by an accretion-powered continuum modified by extinction.
Accordingly, we regard partial envelope dispersal as a possible interpretation rather than a unique explanation of the observed properties.

\acknowledgments
The authors used ChatGPT version 6.1 (OpenAI) to improve the English language, grammar, and clarity of the manuscript. The authors also used Claude-Opus version 5.5 (Anthropic) to assist in the development of the code used to create Figure~\ref{fig:betabeta}. All suggested revisions were reviewed and edited by the authors, who take full responsibility for the final content.
This work is supported by the Japan Society for the Promotion of Science (JSPS) KAKENHI (25K01043, K.~Ichikawa).
K.I. also acknowledges support from the JST FOREST Program, Grant Number JPMJFR2466, which helped make this research possible. 
The data products presented herein were retrieved from the Dawn JWST Archive (DJA). DJA is an initiative of the Cosmic Dawn Center (DAWN), which is funded by the Danish National Research Foundation under grant DNRF140.

\vspace{3mm}
\facilities{JWST(NIRSpec, NIRCam), HST(ACS, WFC3), Subaru(HSC), Spitzer(IRAC, MIPS), Herschel(PACS, SPIRE), ALMA, CXO, VLA}

\vspace{12mm}
\bibliographystyle{aasjournal}
\bibliography{JWSTBBQ}
\end{document}